   \documentclass[reprint,aps]{revtex4-1}
                   \usepackage{dcolumn}

\usepackage{amsmath}
\usepackage{graphicx}
\usepackage{bm}
\usepackage{color}

\newcommand{\GeV}{\makebox{ GeV}}
\newcommand{\beq}{\begin{equation}}
\newcommand{\enq}{\end{equation}}
\newcommand{\beqa}{\begin{eqnarray}}
\newcommand{\beqast}{\begin{eqnarray*}}
\newcommand{\enqa}{\end{eqnarray}}
\newcommand{\enqast}{\end{eqnarray*}}

\def\GeV{\nobreak\,\mbox{GeV}}

\begin{document}

\title{ pp Elastic Scattering at LHC Energies }
\author{A. K. Kohara}
\author{E. Ferreira}
\author{T. Kodama}

\begin{abstract}
Using a unified analytic representation for the elastic scattering
amplitudes of pp scattering valid for all energies above 20 GeV, the behavior 
of observables in the LHC collisions in the range $\sqrt{s}$= 2.76 - 14 TeV is
discussed. After the precise description of $d\sigma/dt$ at 7 TeV, we discuss 
the energy dependence of the amplitudes, and expect that the proposed 
analytical forms  give equally good predictions for  the  future experiments.
\end{abstract}

\maketitle
\affiliation{Instituto de F\'{\i}sica, Universidade Federal do Rio de Janeiro \\
C.P. 68528, Rio de Janeiro 21945-970, RJ, Brazil }
\affiliation{Instituto de F\'{\i}sica, Universidade Federal do Rio de Janeiro \\
C.P. 68528, Rio de Janeiro 21945-970, RJ, Brazil }
\affiliation{Instituto de F\'{\i}sica, Universidade Federal do Rio de Janeiro \\
C.P. 68528, Rio de Janeiro 21945-970, RJ, Brazil }
\affiliation{Instituto de F\'{\i}sica, Universidade Federal do Rio de Janeiro \\
C.P. 68528, Rio de Janeiro 21945-970, RJ, Brazil }
\affiliation{Instituto de F\'{\i}sica, Universidade Federal do Rio de Janeiro \\
C.P. 68528, Rio de Janeiro 21945-970, RJ, Brazil }
\affiliation{Instituto de F\'{\i}sica, Universidade Federal do Rio de Janeiro \\
C.P. 68528, Rio de Janeiro 21945-970, RJ, Brazil }





\section{Introduction\label{intro}}

Elastic scattering is described by one single complex function depending on
two kinetic variables: the incident center of mass energy $\sqrt{s}$ and
momentum transfer $\vec{q}$. In high energy pp(\={p}) scattering, the
scattering amplitude is usually represented as $T(s,t)$, where $t$ is the
four momentum transfer squared. More than a decade ago, Ferreira and Pereira
analyzed all available elastic scattering data for energies above 20 GeV 
\cite{ferreira1} and all $|t|$, identifying properties of the amplitudes
(zeros, signs, magnitudes), with proper attention given to the real part,
which plays a critical role in differential cross sections for mid and large 
$|t|$ ranges.

Recently, this analysis was extended \cite{KEK_2013} to the LHC-TOTEM
elastic scattering 7 TeV data \cite{TOTEM_7},  and also the behavior of
proposed amplitudes was re-examined in the whole energy region from 20 GeV 
to 14 TeV to determine the precise
energy dependence of the model parameters \cite{KEK-to-be}, 
and applied to the cosmic energy domain with calculation of p-air 
cross sections \cite{CR_2014}. From this analysis, an analytic representation 
of scattering amplitudes as function of $\sqrt{s}$ and $t$ was established. 
In the present work we apply these analytical forms
to investigate in detail the LHC energy region from 2.76 to 14 TeV.

We stress that we establish explicitly disentangled real and imaginary
amplitudes based on a QCD motivated model, and not just fit pure
phenomenological expressions to observables. Besides, since the so-called
impact parameter representation   $(s,\vec{b})$     and
its Fourier transform in $(s,\vec{q })$  space are both represented by 
simple analytical forms, 
we are able to control unitarity and dispersion relation constraints, and provide
geometric interpretation of the interaction range. The regularity that we
obtain in our treatment of the data and associated reasonable physical
interpretation of the consequences give reliability to our proposal of
disentanglement of the amplitudes \cite{KEK-to-be}.

The present work is organized as follows. In the next section, we describe
briefly the amplitudes and their energy dependences in both $t$- and $b$-
representations. In Sec. \ref{energy_range}  we apply these amplitude to 
describe observables and discuss their energy dependence, and also we investigate
consequences for very high energies of the form of the amplitudes in $b$%
-space. In Sec. \ref{data} we make use of the properties of our amplitudes and
observables for the LHC range, particularly for $\sqrt{s}=$ 8 TeV, where  
preliminary information on $d\sigma/dt$ starts to become available. 
 The last section is devoted to discuss further our results and
perspectives, together with geometric interpretation in the $b$-space
representation. \bigskip

\section{Analytic representation of the amplitudes}


\subsection{ Impact parameter Representation}

The Fourier transform of the momentum transfer $\vec{q}$ amplitudes to the $%
\vec{b}$-space defines the impact parameter (or simply $b$-space)
representation. Since the impact parameter variable $\vec{b}$ is not
observable, the treatments of data are made usually in $(s,t)$ space, except
for integrated cross sections. However the $b$-space description gives
insight in geometric aspects of the collision, since in the classical limit
the variable $b$ reduces to the physical impact parameter. Besides, it plays
important role in the eikonal representation, where unitarity constraints
are more simply formulated. On the other hand, the dispersion relation (causality)
constraint is properly dealt in $t$-space. In the following discussion, we do not
consider effects of spin or polarization.

 The amplitudes for the description of pp scattering in the
Stochastic Vacuum Model \cite{dosch} are originally constructed 
through prof ile functions in  $b$%
 -space and here we describe this formalism first. The
dimensionless $(s,b)$ amplitudes due to the nuclear interaction  are 
written as 
\begin{equation}
\widetilde{T}_{K}(s,\vec{b})=\frac{\alpha _{K}}{2\beta _{K}}e^{-{b^{2}}/{%
4\beta _{K}}}+\lambda _{K}\widetilde{\psi }_{K}(s,b)~,  \label{b-AmplitudeN}
\end{equation}%
with the characteristic shape function 
\begin{equation}
\widetilde{\psi }_{K}(s,b)=\frac{2e^{\gamma _{K}-\sqrt{\gamma _{K}^{2}+{b^{2}%
}/{a_{0}}}}}{a_{0}\sqrt{\gamma _{K}^{2}+{b^{2}}/{a_{0}}}}\Big[1-e^{\gamma
_{K}-\sqrt{\gamma _{K}^{2}+{b^{2}}/{a_{0}}}}\Big]~.  \label{Shape-b}
\end{equation}%
The label $K$ $=R,I$ indicates either the real or the imaginary part of the
complex amplitude.

The fixed quantity $a_0=1.39 \GeV^{-2}$ is related to the square of the 
correlation length $a$ of the correlation function of the gluon 
condensate, with   $~a~=(0.2\sim0.3)$ fm ~, as measured in hadronic 
interactions and in lattice QCD, with our best choice 0.27 fm.
 In the large $b$ behaviour of the profile function of the Stochastic 
Vacuum Model there appears 
the   dimensionless combination $b^2/a_0$  where 
$a_0 = [a/(3\pi/8)]^2 $ , that fixes the value of $a_0$
  appearing in Eq. (\ref{Shape-b}).  The 
quantity $3\pi/8$ is a feature of the correlation function \cite{dosch} .

The Gaussian form of the first term in Eq. (\ref{b-AmplitudeN}) is similar 
to the usual  formalism of reggeon exchanges  \cite{Regge}. The second term, 
referred to as shape function, represents contributions from the perturbed 
vacuum structure around the protons at larger $b$ values. It is zero at $b=0$ 
and  is  normalized as 
\begin{equation}
\frac{1}{2\pi }\int d^{2}\vec{b}\ \tilde{\psi}_{K}\left( b,s\right) =1~.
\label{psinorm}
\end{equation}%
In Eq. (\ref{b-AmplitudeN}) we have introduced four energy dependent
parameters for each amplitude, $\alpha _{K}$, $\beta _{K}$, $\gamma _{K}$, $%
\lambda _{K}$, with $\gamma _{K}$ dimensionless, while $\alpha _{K}$, $%
\gamma _{K}$ and $\beta _{K}$ are like GeV$^{-2}$. 

In the small and mid $b$ ranges there is superposition of the contributions 
of the two parts, that, for convenience of language, we may call respectively
Regge phenomenology and loop-loop interaction. The resulting parameter values 
are determined describing with accuracy the imaginary and real amplitudes
as a whole, there is no case of double counting effects, and each 
part is duly represented,  if one thinks of each one separately. 
Actually  Eq. (1) represents an extension of parametrization of results  
of the Stochastic Vaccum Model, opening possibilities of introducing 
proper $s$ and $t$ dependences. 

Although $b$ is not exactly the physical impact parameter, neither
observable, the $b$-space representation permits a geometrical
interpretation of the behavior of the amplitude. For large $b$, which
corresponds to peripheral collisions, the amplitudes fall down with a
Yukawa-like tail, 
\begin{equation}
\sim \frac{1}{b}e^{-b/b_{0}},
\end{equation}%
that reflects the effects of virtual partons (the modified gluon field)  at
large distance in the Stochastic Vacuum Model. A feature of the $b$%
-space representation is that it can be directly related the eikonal
formalism, as shown below.

We introduce the eikonal function $\chi \left( s,b\right) $ through 
\begin{equation}
i\sqrt{\pi }~(1-e^{i\chi (s,\vec{b})})~\equiv \widetilde{T}(s,\vec{b})=%
\widetilde{T}_{R}(s,\vec{b})+i\widetilde{T}_{I}(s,\vec{b}),  \label{Eikonal}
\end{equation}%
with 
\begin{equation}
\chi (s,\vec{b})=\chi _{R}(s,\vec{b})+i\chi _{I}(s,\vec{b})~.
\end{equation}%
Separating real and imaginary parts, we have%
\begin{eqnarray}
1-\cos \chi _{R}\ e^{-\chi _{I}} &=&\frac{1}{\sqrt{\pi }}\widetilde{T}_{I}(s,%
\vec{b}),  \label{cosI} \\
\sin \chi _{R}\ e^{-\chi _{I}} &=&\frac{1}{\sqrt{\pi }}\widetilde{T}_{R}(s,%
\vec{b}).  \label{sinR}
\end{eqnarray}%
From Eq. (\ref{sinR}) we have immediately%
\begin{equation}
e^{-2\chi _{I}}\geq \frac{1}{\pi }\widetilde{T}_{R}^{2}(s,\vec{b}),
\end{equation}%
and thus the general unitarity constraint is witten as 
\begin{equation}
\frac{\widetilde{T}_{R}^{2}}{\pi }\leq e^{-2\chi _{I}(s,\vec{b})}\leq ~1~,
\label{bounds}
\end{equation}%
or 
\begin{equation*}
0\leq \chi _{I}\leq -\frac{1}{2}\log (\widetilde{T}_{R}^{2}/\pi )~.
\end{equation*}%
Our solution, at all energies, satisfy this bound condition.

Satisfying a monotonic behavior of the scattering amplitudes, our solutions
are restricted to the branch where $\chi _{R}\geq 0,$ and thus in turn, we
have 
\begin{equation}
0~\leq ~\widetilde{T}_{I}(s,\vec{b})\leq \sqrt{\pi }~~,\ ~\forall ~s,b~.
\label{sector}
\end{equation}%
Under these conditions, our analysis shows that for a fixed $\sqrt{s}$, the
function $\widetilde{T}_{I}(s,\vec{b})$ is monotonically decreasing in $b$.
The maximum of the imaginary amplitude, $\widetilde{T}_{I}(s,\vec{b}=0)$
tends to its limiting value $\sqrt{\pi }$ for asymptotic large energies \cite%
{KEK-to-be}.

In terms of the $\widetilde{T}_{K}(s,\vec{b})$ amplitudes, the elastic,
total and inelastic cross sections are written respectively  
\begin{equation}
\sigma _{\mathrm{el}}(s) =\frac{(\hbar c)^{2}}{\pi }\int d^{2}\vec{b}~|%
\widetilde{T}(s,\vec{b})|^{2} \equiv \int d^{2}\vec{b}~\frac{d\widetilde{%
\sigma }_{\mathrm{el}}(s,\vec{b})}{d^{2}\vec{b}}~,
\end{equation}%
\begin{equation}
\sigma(s) =\frac{2}{\sqrt{\pi }}(\hbar c)^{2}\int d^{2}\vec{b}~\widetilde{T}%
_{I}(s,\vec{b})~ \equiv \int d^{2}\vec{b}~\frac{d\widetilde{\sigma }_{%
\mathrm{tot}}(s,\vec{b})}{d^{2}\vec{b}}~~,
\end{equation}
and 
\begin{eqnarray}
\sigma _{\mathrm{inel}} &=&\sigma-\sigma _{\mathrm{el}} =(\hbar c)^{2}\int
d^{2}\vec{b}~\Bigg(\frac{2}{\sqrt{\pi }}\widetilde{T}_{I}(s,\vec{b})-\frac{1%
}{\pi }|\widetilde{T}(s,\vec{b})|^{2}\Bigg)  \notag \\
&\equiv &\int d^{2}\vec{b}~\frac{d\widetilde{\sigma }_{\mathrm{inel}}(s,\vec{%
b})}{d^{2}\vec{b}}~.  \label{dsdb_inel}
\end{eqnarray}%
In terms of the eikonal function, we write%
\begin{eqnarray}
\frac{d\widetilde{\sigma }_{\mathrm{el}}(s,\vec{b})}{d^{2}\vec{b}}
&=&1-2\cos \chi _{R}e^{-\chi _{I}}+e^{-2\chi _{I}}, \\
\frac{d\widetilde{\sigma }(s,\vec{b})}{d^{2}\vec{b}} &=&2\left( 1-\cos \chi
_{R}e^{-\chi _{I}}\right) \\
\frac{d\widetilde{\sigma }_{\mathrm{inel}}(s,\vec{b})}{d^{2}\vec{b}}
&=&1-e^{-2\chi _{I}}.
\end{eqnarray}

\subsection{ $t$-space representation}

The comparison with $d\sigma /dt$ data and determination of parameters 
are made with the amplitudes in $t$-space. The quantities $\Psi
_{K}(\gamma _{K}(s),t=-\vec{q}_{T}^{2})$ obtained by Fourier transform of
Eq. (\ref{b-AmplitudeN}) are written

\begin{equation}
T_{K}^{N}(s,t)=\alpha _{K}(s)\mathrm{e}^{-\beta _{K}(s)|t|}+\lambda
_{K}(s)\Psi _{K}(\gamma _{K}(s),t),  \label{TN-b1}
\end{equation}%
with $K=R,I$, and the shape functions in $t-$ space take the form 
\begin{eqnarray}
&&\Psi _{K}(\gamma _{K}(s),t) \\
&=&2~\mathrm{e}^{\gamma _{K}}~\bigg[{\frac{\mathrm{e}^{-\gamma _{K}\sqrt{%
1+a_{0}|t|}}}{\sqrt{1+a_{0}|t|}}}-\mathrm{e}^{\gamma _{K}}~{\frac{e^{-\gamma
_{K}\sqrt{4+a_{0}|t|}}}{\sqrt{4+a_{0}|t|}}}\bigg]~,  \notag  \label{psi_st}
\end{eqnarray}%
%
with the property 
\begin{equation}
\Psi _{K}(\gamma _{K}(s),t=0)=1~,  \label{psinorm2}
\end{equation}%
that corresponds to Eq. (\ref{psinorm}).

The expression (\ref{TN-b1}) represents the nuclear amplitude due to the
non-perturbative QCD interactions that dominate the low and mid $|t|$
regions. To describe elastic $d\sigma /dt$ data for all $|t|$, we should
account for contributions from perturbative processes. We thus add a term
representing the perturbative three-gluon exchange amplitude \cite{DL} that
may appear in the large $|t|$ region, and the complete nuclear amplitudes
are then written 
\begin{eqnarray}
&& T_{K}^{N}(s,t)\rightarrow T_{K}^{N}(s,t) \\
&& =\alpha _{K}(s)\mathrm{e}^{-\beta_{K}(s)|t|}+ \lambda _{K}(s)\Psi
_{K}(\gamma _{K}(s),t)  \notag \\
&& +\delta_{K,R}R_{ggg}\left( t\right) ,~~~K=R,I~,  \notag
\label{hadronic_complete}
\end{eqnarray}
where the Kronecker delta symbol $\delta _{K,R}$ is introduced since we
define $R_{ggg}\left( t\right) $ as the real contribution from the
perturbative three-gluon exchange amplitude. The effect of the tail term $%
R_{ggg}\left( t\right) $, producing a universal (not energy dependent) $%
|t|^{-8}$ form for large $|t|$ in $d\sigma/dt$, was studied in the analysis
of the experiments at CERN-ISR, CERN-SPS \cite{ferreira1}, 1.8 TeV \cite%
{KEK_2013b} and 7 TeV \cite{KEK_2013}.
We write  
\begin{equation}
R_{ggg}(t)\equiv\pm0.45~t^{-4}(1-e^{-0.005\left\vert t\right\vert ^{4}%
})(1-e^{-0.1\left\vert t\right\vert ^{2}})~,\label{R_tail}%
\end{equation}
where the last two factors cut-off  this term smoothly in the non-perturbative
domain, and the signs $\pm$ refer to the pp and p$\mathrm{{\bar{p}}}$
amplitudes respectively.
Although the cut-off factors written in Eq. (\ref{R_tail}) have been adequate 
for all cases that were examined, their detailed forms in the transition range 
$(2.5~<~|t|~<~4)$ ~ GeV$^{2}$  must be examined with data.

For a complete analysis of elastic scattering, we must also take into
account the contribution from the Coulomb interaction. The complete
amplitudes $T_{R}(s,t)$ and $T_{I}(s,t)$, with dimensions GeV$^{-2}$,
contain the nuclear and the Coulomb parts as 
\begin{equation}
T_{R}(s,t)=T_{R}^{N}(s,t)+\sqrt{\pi }F^{C}(t)\cos (\alpha \Phi )~,
\label{real}
\end{equation}%
and 
\begin{equation}
T_{I}(s,t)=T_{I}^{N}(s,t)+\sqrt{\pi }F^{C}(t)\sin (\alpha \Phi )~,
\label{imag}
\end{equation}%
where $\alpha ~$is the fine-structure constant, $\Phi (s,t)$ is the Coulomb
phase and $F^{C}(t)$ is related with the proton form factor 
\begin{equation}
F^{C}(t)~=(-/+)~\frac{2\alpha }{|t|}~F_{\mathrm{proton}}^{2}(t)~,
\label{coulomb}
\end{equation}%
for the pp$/$p$\mathrm{{\bar{p}}}$ collisions. The proton form factor is
taken as%
\begin{equation}
F_{\mathrm{proton}}(t)=[t_{0}/(t_{0}+|t|)]^{2}~,  \label{ff_proton}
\end{equation}%
where $t_{0}=0.71\ $GeV$^{2}$. Note that the strong interaction part of the
amplitudes are smooth and regular functions of $s$ and $t$, while the
Coulomb amplitude is relevant in the very forward range $|t|<10^{-2}~\mathrm{%
{GeV}^{2}}$.

In our normalization the elastic differential cross section is written 
\begin{eqnarray}  \label{Sigma_diff}
\frac{d\sigma(s,t)}{dt}&=& (\hbar c)^2[T_I^2(s,t)+ T_R^2(s,t)] \\
&=& \frac{d\sigma^I(s,t)}{dt} + \frac{d\sigma^R(s,t)}{dt} ~ ,  \notag
\end{eqnarray}
and the total pp cross section is given by the optical theorem 
\begin{eqnarray}  \label{Sigma_total}
\sigma = (\hbar c)^2~ 4\sqrt{\pi} ~ T_{I}^{N}(s,t=0)~.
\end{eqnarray}

The analysis of all pp elastic scattering data for $\sqrt{s}$ from 20 GeV to
7 TeV leads to a separate identification of the real and imaginary parts
contributing to Eq. (\ref{Sigma_diff}). The energy dependence of the eight
parameters is given below, with $\sqrt{s}$ in TeV, and $\nobreak\,\mbox{GeV}%
^{-2}$ in the units of the parameters that are not dimensionless ( $\gamma_I$
and $\gamma_R$ are dimensionless). 
\begin{equation}  \label{a1}
\alpha_I(s)=11.0935+1.35479\log\sqrt{s} ,
\end{equation}
\begin{eqnarray}  \label{a2}
\beta_I(s)=&& 4.44606586+0.3208411\log\Big(\sqrt{s}/{30.4469}\Big)  \notag \\
&&+0.0613381 \Big[\log^2\Big(\sqrt{s}/30.4469\Big)+0.5 \Big]^{1/2},
\end{eqnarray}
\begin{equation}  \label{a7}
\alpha_R(s)=0.208528+0.0419028~\log\sqrt{s} ~ ,
\end{equation}
\begin{equation}  \label{a8}
\beta_R(s)=1.1506+0.12584~\log\sqrt{s}+ 0.017002~\log^2\sqrt{s} ~ ,
\end{equation}
\begin{equation}  \label{a5}
\gamma_I(s)=10.025+0.79097~\log\sqrt{s}+0.088~\log^2\sqrt{s} ~ ,
\end{equation}
\begin{equation}  \label{a6}
\gamma_R(s)=10.401+1.4408~\log(\sqrt{s})+0.16659~\log^2(\sqrt{s}) ~ ,
\end{equation}
\begin{equation}  \label{a3}
\lambda_I(s)=14.02008+3.23842~\log\sqrt{s}+0.444594~\log^2\sqrt{s} ~ ,
\end{equation}
\begin{equation}  \label{a4}
\lambda_R(s)=3.31949+0.743706~\log\sqrt{s} ~.
\end{equation}

The peculiar (not so simple) expression for $\beta _{I}(s)$ is constructed
in order to satisfy both the low-energy phenomenology and unitarity
constraints at all energies, as given in Eq. (\ref{sector}), and leads to
the asymptotic behavior $\widetilde T_I(s,b=0) \rightarrow \sqrt{\pi}$.
 For very high energy and considerations of asymptotic behaviour, 
it is useful to  use the simpler form for $\beta_I(s)$
\begin{equation}
\label{a2_HE}
\beta_I(s) = 0.382179  \log(\sqrt{s})+ 3.14055   
\end{equation}

The first term in Eq. (\ref{TN-b1}) can be written in the usual notation  
of Regge phenomenology, with the dimensionless scattering amplitude $A(s,t)$ 
\begin{eqnarray}
&& A(s,t) ~ > \rightarrow [4 \sqrt{\pi}\times 11.09\times 10^{6} {\rm e}^{-3.14|t|}] \\
&& \times [1+0.061 \log(s/1 {\rm TeV^2})] \times (s/1 {\rm TeV^2})^{1-0.19 |t|} ~ , \nonumber
\end{eqnarray} 
where $|t|$ is in $\GeV^2$ and $\sqrt{s}$ in TeV, with a 
$t$-dependent residue and a trajectory with intercept 1 and angular coefficient 
0.19 GeV$^{-2}$. The log term corresponds to a double pole, arising from derivative 
with respect to the trajectory \cite{Regge}.  
 
 These expressions are able to give high precision representation for all 
data \cite{ferreira1,KEK-to-be,KEK_2013,KEK_2013b}, with coherent and reliable 
identification of the  real and imaginary   amplitudes.  
Properties and consequences for
the energy range above 1 TeV are discussed in the present paper, with
particular attention to the experimental LHC energies. Cosmic ray energies
up to $\sqrt{s}$ =100 TeV and asymptotic behaviour have been discussed 
elsewhere \cite{CR_2014}. 

\subsection{Forward Amplitudes and Associated Observables\label{forward}}

In the very forward direction, where the elastic pp and p$\mathrm{{\bar{p}}}$
scattering amplitudes can be approximated by pure exponential forms, the
differential cross section is written 
\begin{eqnarray}
\frac{d\sigma }{dt} &\rightarrow &\pi \left( \hbar c\right) ^{2}~~\Big\{\Big[%
\frac{\rho \sigma }{4\pi \left( \hbar c\right) ^{2}}~{{e}%
^{B_{R}t/2}+F^{C}(t)\cos {(\alpha \Phi )}\Big]^{2}}  \notag \\
&&+\Big[\frac{\sigma }{4\pi \left( \hbar c\right) ^{2}}~{{e}%
^{B_{I}t/2}+F^{C}(t)\sin {(\alpha \Phi )}\Big]^{2}\Big\}~,}
\label{diffcross_eq}
\end{eqnarray}%
where $t\equiv -|t|$ and we must allow different values for the slopes $B_{I}
$ and $B_{R}$ of the imaginary and real amplitudes. With $\sigma $ in
milibarns and $|t|$ in GeV$^{2}$, we have $\left( \hbar c\right)
^{2}~=~0.3894$. Since we work with $B_{R}\neq B_{I}$ , treatment of the
Coulomb interference requires a more general expression for the Coulomb
phase, which has been developed before \cite{KEK_2013}.

The limits of the amplitudes for small $|t|$ give the total cross section $%
\sigma$, the ratio $\rho$ of the real to imaginary amplitudes, and the
slopes $B_{R,I}$ at $t=0$ through 
\begin{equation}
\sigma(s)=4\sqrt{\pi} \left(\hbar c\right)^{2}~[\alpha_{I}(s)+\lambda
_{I}(s)]~,  \label{sigma_par}
\end{equation}%
\begin{equation}
\rho(s)=\frac{T_{R}^{N}(s,t=0)}{T_{I}^{N}(s,t=0)}=\frac{\alpha_{R}(s)+%
\lambda_{R}(s)}{\alpha_{I}(s)+\lambda_{I}(s)}~,  \label{rho_par}
\end{equation}
\begin{eqnarray}
B_{K}(s) && =\frac{2}{T_{K}^{N}(s,t)}\frac{dT_{K}^{N}(s,t)}{dt}\Big|_{t=0}=~%
\frac{2}{\alpha_{K}(s)+\lambda_{K}(s)}\times  \notag \\
&&\Big[\alpha_{K}(s)\beta_{K}(s)+\frac{1}{8}\lambda_{K}(s)a_{0}\Big(6\gamma
_{K}(s)+7\Big)\Big]~ .  \label{slopes_par}
\end{eqnarray}

\begin{figure}[b]
\includegraphics[width=8cm]{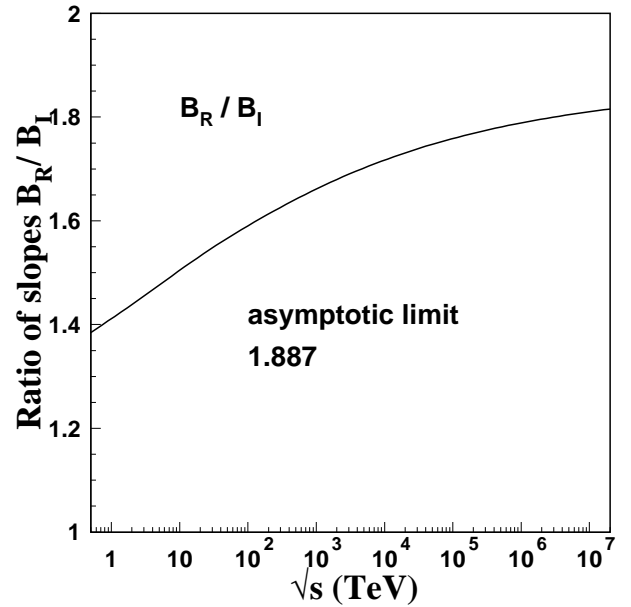} 
\caption{ The slopes of real and imaginary amplitudes vary with the energy
with a $\log ^{2}$ dependence as given by Eqs. (\protect\ref{BI-eq},\protect
\ref{BR-eq}). At all energies it is $B_{R}>B_{I}$, as predicted by
dispersion relations \protect\cite{ferreira2}. In the figure, the ratio $%
B_{R}/B_{I}$ is plotted as function of the energy, indicating the finite
asymptotic limit.}
\label{BRBIratio-fig}
\end{figure}

Using the energy dependences given in Eqs. (\ref{a1}-\ref{a4}) 
we can write the practical expressions for the four quantities 
\begin{equation}
\sigma (s)=69.3286+12.6800\log \sqrt{s}+1.2273\log ^{2}\sqrt{s}\ ,
\label{sig-eq}
\end{equation}%
\begin{equation}
B_{I}(s)=16.2472+1.53921\log \sqrt{s}+0.174759\log ^{2}\sqrt{s}\ ,
\label{BI-eq}
\end{equation}%
\begin{equation}
B_{R}(s)=22.835+2.862\log \sqrt{s}+0.329721\log ^{2}\sqrt{s}\ ,
\label{BR-eq}
\end{equation}%
and 
\begin{equation}
\rho (s)=\frac{3.528018+0.7856088\log \sqrt{s}}{25.11358+4.59321\log \sqrt{s}%
+0.444594\log ^{2}\sqrt{s}}~,  \label{rho-eq}
\end{equation}%
where $\sqrt{s}$ is in TeV, $\sigma $ in milibarns, $B_{I}$ and $B_{R}$ are
in $\nobreak\,\mbox{GeV}^{-2}$; $\rho $ is dimensionless, passes through a
maximum at about 1.8 TeV, and decreases at higher energies, with asymptotic
value zero. The ratio $B_{R}/B_{I}$ is always larger than one, as expected
from dispersion relations \cite{ferreira2}. The ratio $B_{R}/B_{I}$ as
function of the energy is shown in Fig. (\ref{BRBIratio-fig}). There is a
finite asymptotic value $B_{R}/B_{I}\rightarrow 1.887$. 

This treatment of pp  forward scattering has been applied to the calculation 
of p-air cross sections measured in Extend Air Showers studies in cosmic 
ray experiments. Covering the range from 1 to 100 TeV  in pp energies,
our  input amplitudes are used as basis of Glauber calculations, 
 giving  good description \cite{CR_2014}  of all cosmic ray data.  

\clearpage

\section{Observables in the range from 1.8 to 14 TeV \label{energy_range}}

\subsection{Differential Cross Sections and Amplitudes}

\begin{figure}[b]
\includegraphics[width=8cm]{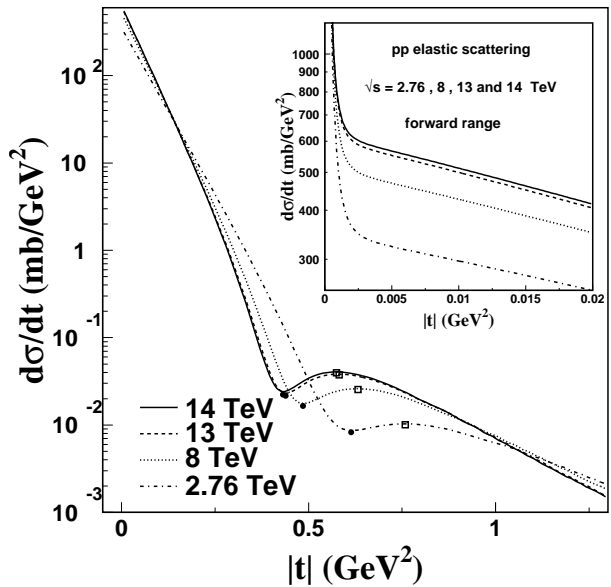}
\caption{ The lines show the values of $d\protect\sigma /dt$ obtained for
energies of LHC experiments. The 7 TeV case, presented before \protect\cite%
{KEK_2013}, is obviously very close to the 8 TeV curve. The positions of
dips and bump peaks at different energies, marked with dots and squares, can
be connected with straight lines. The inset shows the low $|t|$ range, with
Coulomb interaction effects included. }
\label{pilhas-fig}
\end{figure}

In Fig. \ref{pilhas-fig} we show the predictions for $d\sigma /dt$ for the
LHC energies 2.76 , 8 , 13 and 14 TeV. We first observe that the dip and the
bump peak displace to the left as the energy increases and in this figure
these displacements follow almost straight lines, as indicated by marks with
black circles and open squares. For the sake of convenience, we list the
values of parameters for these energies in Table \ref{first_table},  where $%
\gamma _{I},\lambda _{I},\alpha _{R}$ and $\gamma _{R}$ are substituted by
more commonly used quantities $\sigma,\rho \ $ together with the slope
parameters $B_{I}$ and $B_{R}$. In Table \ref{second_table} we show the
values of several quantities obtained in the numerical calculation of the
amplitudes and of observables in the elastic process. Some characteristic
features are exhibited below in plots. 
\begin{table*}[tbp]
\caption{ Values of parameters that build the amplitudes for all $|t|$, for
the energies of LHC pp collisions. \protect\vspace{0.1cm} }
\label{first_table}\tabcolsep=0.009cm 
\begin{tabular}{|c|cccc|cccc|}
\hline
& \multicolumn{4}{c|}{imaginary amplitude} & \multicolumn{4}{c|}{real
amplitude} \\ \cline{2-9}
$\sqrt{s}$ & $\sigma $ & $B_I$ & $\alpha_I$ & ~$\beta_{I} $ ~~ & $\rho$ & $%
B_R$ & $\lambda_{R} $ & $\beta_R$ \\ 
TeV & mb & GeV$^{-2}$ & GeV$^{-2}$ & GeV$^{-2}$ &  & GeV$^{-2}$ & GeV$^{-2}$
& GeV$^{-2}$ \\ \hline
1.8 & ~ 77.21 ~ & ~ 17.17 ~ & ~ 11.8898 ~ & ~ 3.7175 ~ & ~ 0.1427 ~ & ~
24.63 ~ & ~ 3.7566 ~ & ~ 1.2304 ~ \\ 
2.76 & 83.47 & 17.96 & 12.4689 & 3.8293 & 0.1431 & 26.08 & 4.0745 & 1.2959
\\ 
7 & ~ 98.65 ~ & ~ 19.90 ~ & ~ 13.7298 ~ & ~ 4.0745 ~ & ~ 0.1415 ~ & ~ 29.65 ~
& ~ 4.7667 ~ & ~ 1.4599 ~ \\ 
8 & 101.00 & 20.21 & 13.9107 & 4.1100 & 0.1411 & 30.21 & 4.8660 & 1.4858 \\ 
13 & 109.93 & 21.35 & 14.5685 & 4.2409 & 0.1392 & 32.35 & 5.2271 & 1.5852 \\ 
14 & 111.34~ & ~ 21.53 ~ & ~ 14.6689 ~ & ~ 4.2612 ~ & ~ 0.1389~ & ~ 32.68~
& ~ 5.2822~ & ~ 1.6011 ~ \\ \hline
\end{tabular}%
\end{table*}
\begin{table*}[tbp]
\caption{ Some derived quantities   that characterize the structure of
amplitudes and cross sections : positions of zeros, dip, and $|t|_{\mathrm{%
peak}} $ at highest point of bump in $d\protect\sigma /dt$ ; ratio R of
values of $d\protect\sigma /dt$ at $|t|_{\mathrm{peak}}$ and $|t|_{\mathrm{%
dip}}$ ; position and height of  the inflection ; 
inelastic and integrated elastic cross sections. \protect\vspace{%
0.1cm} }
\label{second_table}
\begin{tabular}{|c|ccccccccccccccc|}
\hline
$\sqrt{s}$ & $\mathrm{Z_{I}} $ ~ & $\mathrm{Z_{R}(1)} $~ & $\mathrm{Z_{R}(2) 
}$~ & $|t|_{\mathrm{dip}}$ & $d\sigma/dt|_{\mathrm{dip}} $ & $|t|_{\mathrm{%
peak}}$ & $d\sigma/dt|_{\mathrm{peak}} $ & ratio & $|t|_{\mathrm{
infl}}$ & $d\sigma/dt|_{\mathrm{infl}} $    &   $\sigma_{\mathrm{inel}}$
& $\sigma_{\mathrm{el}}$ & $\sigma_{\mathrm{el}}^I$ & $\sigma_{\mathrm{el}%
}^R $ & $\sigma_{\mathrm{el}}/\sigma$ \\ 
TeV & GeV$^{2}$ & GeV$^{2} $ & GeV$^{2}$ & GeV$^{2}$ & mb/GeV$^{2}$ & GeV$%
^{2}$ & mb/GeV$^{2}$ & R &  GeV$^{2}$   & mb/GeV$^{2}$     &  mb & mb & mb & mb &  \\ \hline
1.8 & 0.6250 & 0.2052 & 1.0464 & 0.6798 & 0.00583 & 0.8170 & 0.00663 & 
1.1362 & 0.7289 & 0.00615  & 58.89 & 18.31 & 18.07 & 0.24 & 0.237 \\ \hline
2.76 & 0.5723 & 0.1925 & 0.9788 & 0.6138 & 0.00825 & 0.7587 & 0.01009 & 
1.2221 &0.6633 &0.00896  & 63.11 & 20.35 & 20.09 & 0.27 & 0.244 \\ \hline
7 & 0.4757 & 0.1673 & 0.8445 & 0.4989 & 0.01535 & 0.6465 & 0.02286 & 1.4891& 0.5459&0.01812  &  73.26 & 25.39 & 25.07 & 0.32 & 0.257 \\ \hline
8 & 0.4635 & 0.1639 & 0.8267 & 0.4850 & 0.01659 & 0.6319 & 0.02549 & 1.5368& 0.5314&0.01985 & 74.82 & 26.18 & 25.86 & 0.33 & 0.259 \\ \hline
13 & 0.4225 & 0.1522 & 0.7654 & 0.4385 & 0.02158 & 0.5816 & 0.03742 & 
1.7338 &0.4827  & 0.02732 &  80.79 & 29.20 & 28.85 & 0.35 & 0.266 \\ \hline
14 & 0.4166 & 0.1505 & 0.7565 & 0.4319 & 0.02242 & 0.5743 & 0.03963 & 
1.7678 &0.4758 &  0.02864 & 81.66 & 29.68 & 29.32 & 0.35 & 0.267 \\ \hline
\end{tabular}
\end{table*}
\begin{table*}[tbp]
\caption{ Characteristic values of   $b$-space amplitudes and eikonal functions. 
These quantities are related to the saturation of unitarity bounds.  
Thus $ \widetilde{T}_I(b=0) $ approaches the bound $\sqrt{\pi}=1.77$ as $\sqrt{s}$
increases.    }
\label{third_table}
\begin{tabular}{|c|cccc|}
\hline
$\sqrt{s}$ & $\widetilde{T}_{I}(b=0)  $  & $\widetilde{T}_{R}(b=0)$ & $\chi_{I}(b=0)$ & $\chi_{R}(b=0)$ \\ 
 TeV &   &   &  &  \\ \hline
1.8 & 1.5992   & 0.0947 & 2.1945  &0.5004 \\   \hline
2.76 & 1.6281 & 0.0969 &2.3219  & 0.5910  \\   \hline
7 & 1.6849 & 0.0993  & 2.5939  & 0.8482  \\   \hline
 8 & 1.6923  &0.0995  &2.6299  & 0.8927  \\   \hline
13 &1.7176 &0.0997 &2.7460  & 1.0678  \\   \hline 
14 & 1.7212  & 0.0997 &2.7611  & 1.0958 \\   \hline
\end{tabular}
\end{table*}
In Fig. \ref{amplitudes-fig} we use the energy $\sqrt{s}=8$ TeV as an
example to show the imaginary and real amplitudes $T_I^N(s,t)$, $T_R^N(s,t)$
as functions of $|t|$ as predicted by Eq.(\ref{hadronic_complete}). For all
energies the characteristic features are the two zeros of the real part, and
the single zero of the imaginary part appearing in the plotted range (a
second zero of $T_I^N$ would appear in a much larger $|t|$, outside
experimental visibility). The interplay of the imaginary and real amplitudes
at mid values of $|t|$ is responsible for the dip-bump structure of the
differential cross section, that was shown before \cite{KEK_2013} for $\sqrt{%
s}=7$ TeV, and is exemplified for 8 TeV in the next section. For $|t|\geq
1.5 \nobreak\,\mbox{GeV}^2$ the real part becomes dominant, with positive
sign. The inset shows the small $|t|$ range, in log scale, normalized to one
at $|t|=0$. The straight exponential slopes are shown in dashed lines, with
the dramatic difference between the real and imaginary amplitudes. Soon the
exact amplitudes leave the straight line and curve down, searching for their
respective zeros. As shown in the next section, the consequences for the 
behavior of $d\sigma/dt$ at 8 TeV  will be visible for $|t|$ larger than 
about 0.2 GeV$^2$.

The difference in slopes $B_{R}$ and $B_{I}$ that is required by dispersion
relations \cite{ferreira2}, is often neglected. The real part is small for
small $|t|$, due to the small value of $\rho $, but becomes influential or
dominant for mid and large $|t|$. The amplitudes must be treated as
functions for the whole $|t|$ range. Our unique analytical form connects all
regions and controls the behavior both at small and large $|t|$. Thus, for
example, the value of $\rho $ is very important for the shape of the dip-bump 
structure.

\begin{figure}[b]
\includegraphics[width=8cm]{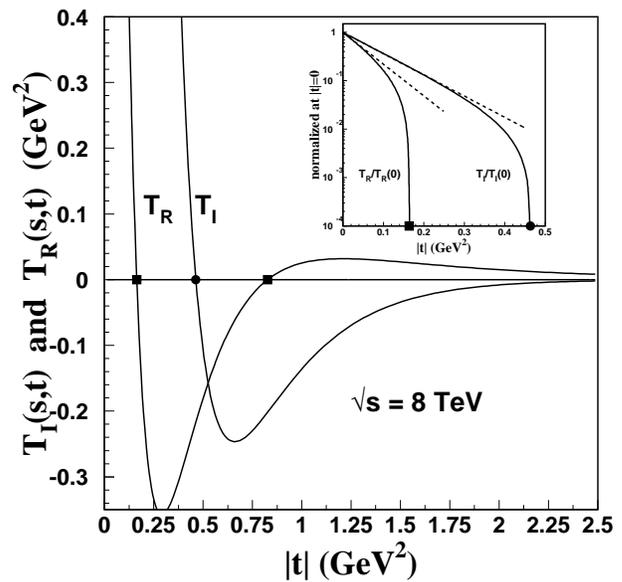} 
\label{amplitudes-fig}
\caption{ Plots of the real and imaginary parts of elastic pp scattering
amplitude at 8 TeV, as functions of $|t|$. The general behaviour is the same
for all energies, with one and two zeros respectively for the imaginary and
real parts. The behaviour for small $|t|$ is shown in the inset, indicating
the difference of slopes $B_R$ and $B_I$ at the origin, and the deviations
of the exponential forms that occur as $|t|$ increases, each amplitude going
towards its zero. A second zero of the imaginary part occurs at much higher $%
|t|$. }
\label{amplitudes-fig}
\end{figure}

\begin{figure*}[b]
\includegraphics[width=8cm]{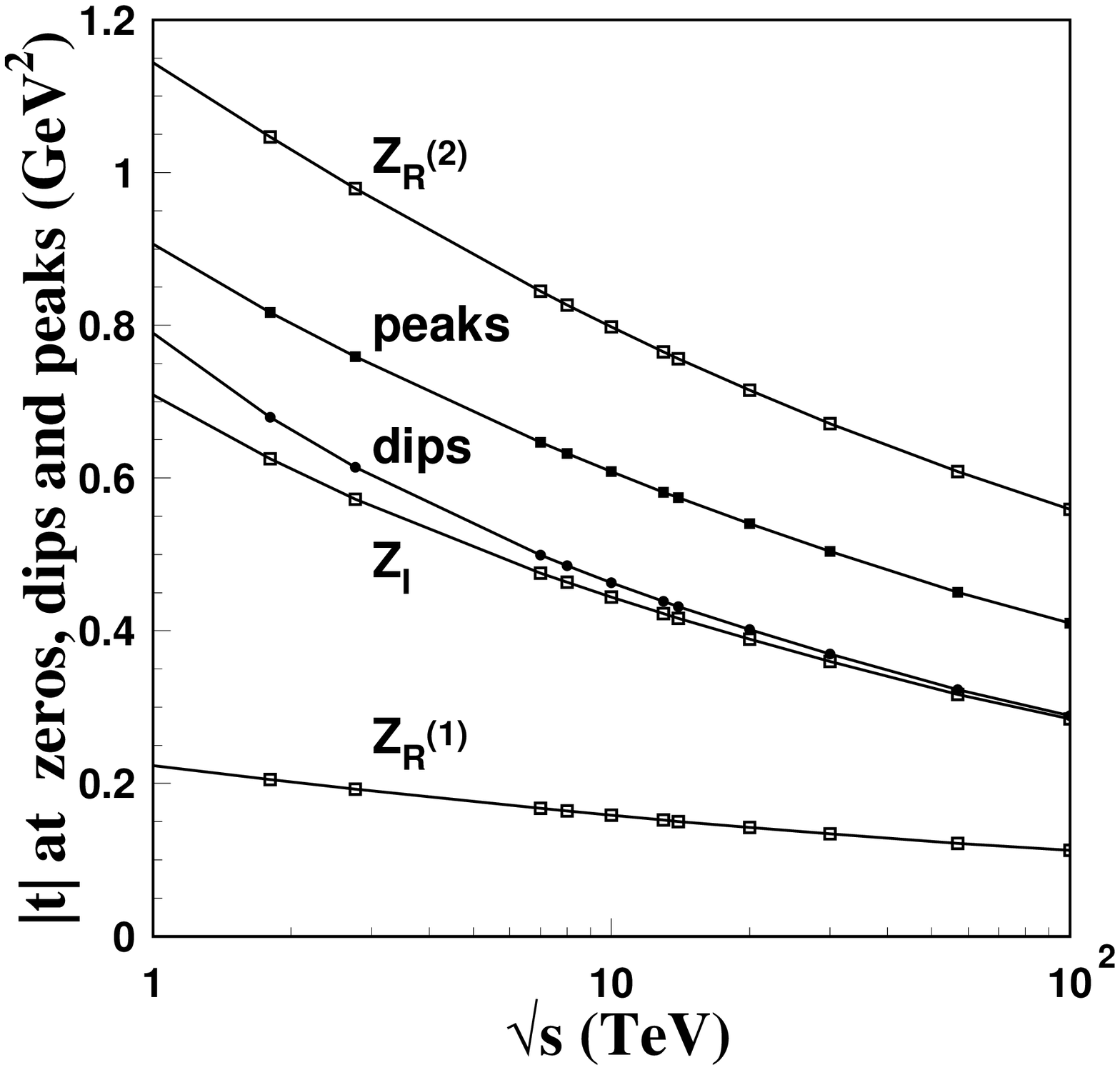} %
\includegraphics[width=8cm]{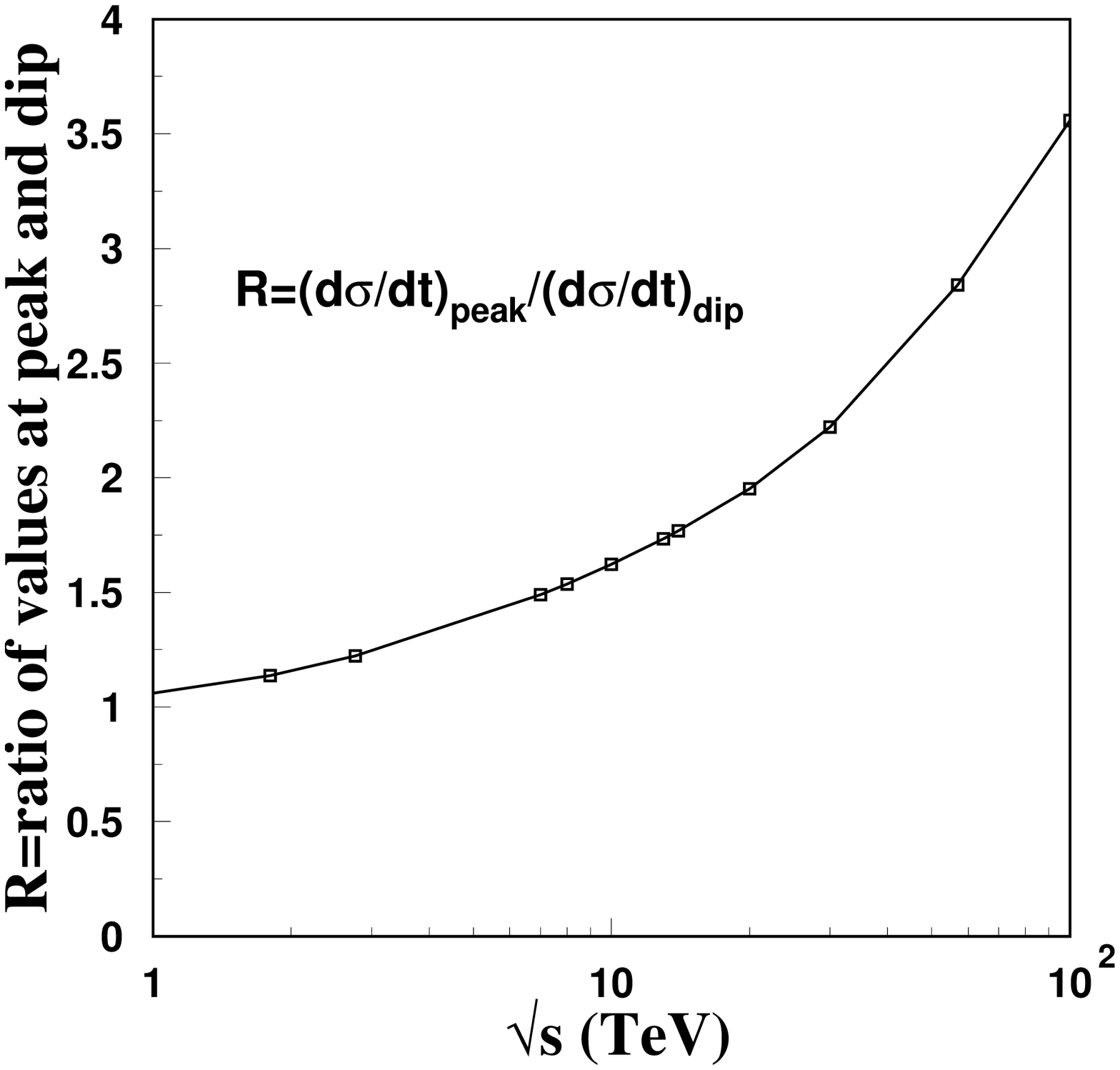}
\caption{ a) Positions of the zeros of the amplitudes, and of the dip and
peak at the bump of $d\protect\sigma/dt$. There appears one zero in the
imaginary and two in the real amplitude. A second imaginary zero occurring
at very large $|t|$ is outside the physically accessible range. All
quantities move towards small values with increasing energies. The dips tend
to coincide with the imaginary zero at high energies. The remarkable
dip/bump structure in pp scattering occurs in the interval between the
imaginary zero and the second real zero. The first real and the imaginary
zero move towards smaller $|t|$, indicating the $\log^2\protect\sqrt s$
increase of the real and imaginary slopes. The dots are put to help the
connection of values of the quantities for different energies. b) There is a
regular and fast increase of the ratio $R=[d\protect\sigma/dt]\mathrm{peak}%
/[d\protect\sigma/dt]\mathrm{dip}$, with increasing sharpness of the
dip/bump structure although the distance $|t|_{\mathrm{peak}}-|t|_{\mathrm{%
dip}}$ between them varies very little. These symptoms come from the
increasing proximity of $|t|_{\mathrm{dip}} $ and $Z_I$, and to the
convergence to finite asymptotic limits of both $|t|_{\mathrm{peak}}$ and $%
|t|_{\mathrm{dip}}$. }
\label{zeros-dips}
\end{figure*}

The regular energy dependence of the positions of the zeros and of dips and
peaks of bumps is shown in Fig. \ref{zeros-dips}. We see that all these
characteristic quantities move towards smaller $|t|$ with increasing energy,
following forms like 
\begin{equation}  \label{martin_theor}
A+\frac{1}{a+b \log\sqrt{s}+c \log^2\sqrt{s} } ~,
\end{equation}
possibly with finite asymptotic limits $A$. Particularly interesting is the
displacement of the first real zero $Z_R^{(1)}$, that at very high energies
behaves as above, with $A=0$ and $c=0$, according to a theorem by A. Martin 
\cite{Martin}. This behaviour is obviously connected with a fast increase of
the slope $B_R$.

It is interesting to observe the relative positions of the dip and the peak
of the bump in $d\sigma/dt$ and the zeros of the imaginary and real parts,
shown in Fig. \ref{zeros-dips}. This question has been discussed a long time
ago \cite{ferreira1}. The figure shows that $Z_I$ and the dip position tend
to the (apparently) common finite limit. Dips and peaks are always located
between $Z_I$ and $Z_R^{(2)}$. All energy dependences are simple and can be
easily parameterized.

It is interesting to note that the ratio between the maximum of the mid-$|t|$
bump (called peak) and the dip minimum 
\begin{equation}
R=[d\sigma /dt]_{\mathrm{peak}}/[d\sigma /dt]_{\mathrm{dip}}
\label{ratiobumpdip}
\end{equation}%
increases with energy rather rapidly (see Fig. \ref{zeros-dips}-b), like $%
\sim \ln ^{2}\sqrt{s}$, while the distance $|t|_{\mathrm{peak}}-|t|_{\mathrm{%
dip}} $ remains practically constant (Fig. \ref{zeros-dips}-a).

\begin{figure}[b]
\includegraphics[width=8cm]{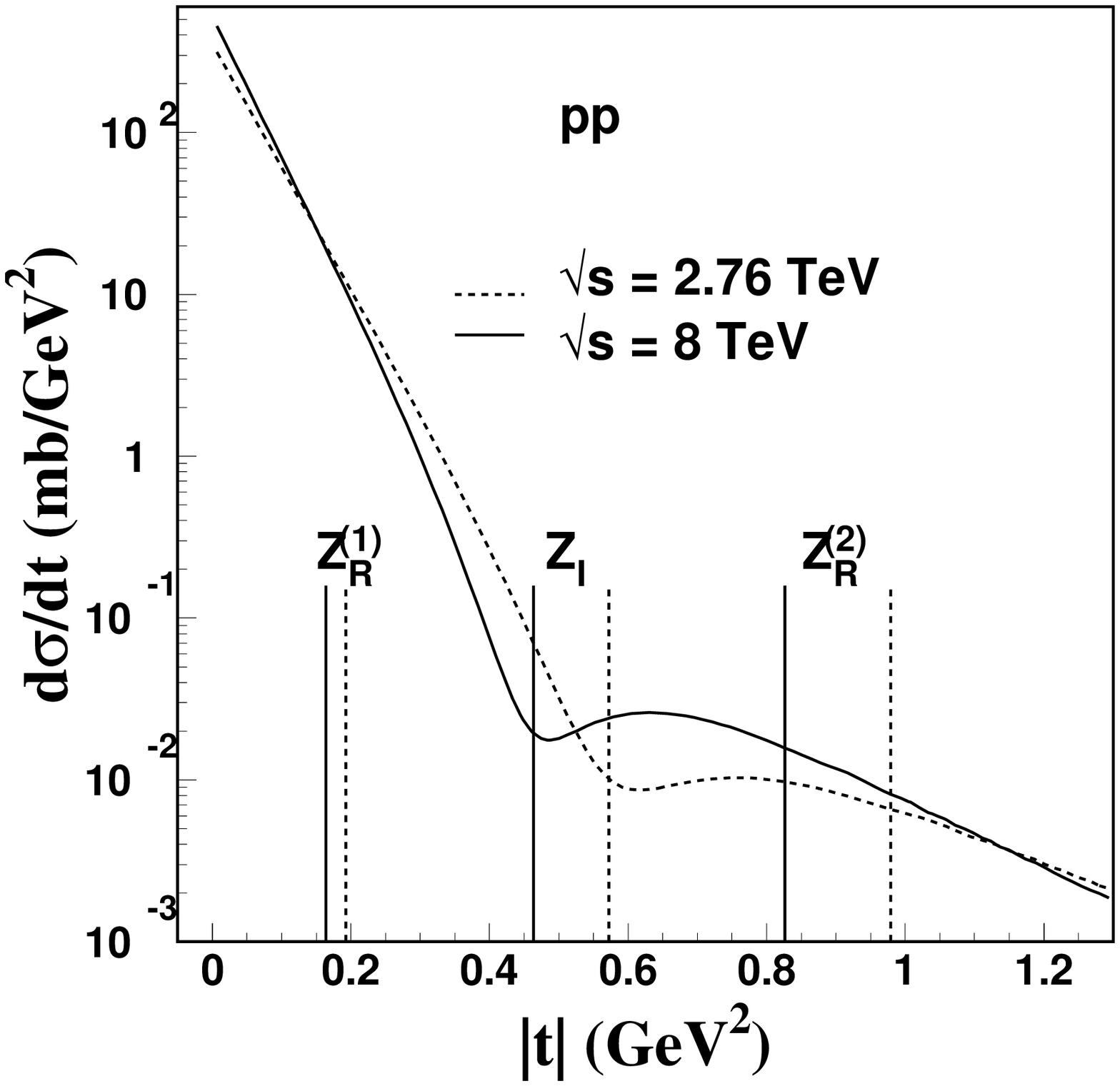}
\caption{ The dip-bump structure in the differential cross section is
determined by the interplay of the regularly increasing modulus (magnitude)
of the imaginary part and the regularly decreasing modulus (magnitude) of
the real part. At all energies both dip and peak of the bump are located
between $Z_I$ and $Z_R^{(2)}$. This behavior is shown in this figure for the
energies 2.76 and 8 TeV. As the energy increases $|t|_{\mathrm{dip}}$
approaches $Z_I$ from the right to the left. Fig. \protect\ref{zeros-dips}
illustrates these properties again, in another way. }
\label{structure-fig}
\end{figure}

In Fig. \ref{structure-fig} we plot $d\sigma/dt$ for 2.76 and 8 TeV, showing
that the characteristic dip/bump structure of $d\sigma/dt$ occurs in the
interval between the imaginary zero and the second real zero. 

\subsection{Integrated Quantities, Ratios and Asymptotic Limits}

The integrated elastic cross section due to the imaginary amplitude can be
represented by 
\begin{eqnarray}
&&\sigma _{\mathrm{el}}^{I}(s)=\int_{0}^{\infty }dt~T_{I}(s,t)^{2}~dt  \notag
\\
&=&15.3366+4.15903\log \sqrt{s}+0.43405\log ^{2}\sqrt{s}~,
\label{elasticI_param}
\end{eqnarray}%
with $\sqrt{s}$ in TeV and $\sigma _{\mathrm{el}}^{I}(s)$ in mb. The
accuracy of this representation is very good, particularly for energies
equal and above 7 TeV . The ratio with the total cross section has a finite
asymptotic limit at high energies $\sigma _{\mathrm{el}}^{I}/\sigma
\rightarrow 0.354$. This result is very important for a geometrical
description of pp scattering, as it means that pp collision does not follows
a black disk form at high energies (see below).

For the contribution of the real part to the elastic cross section the
quantity that is related to the exponential behaviour in the forward
direction and that presents a finite asymptotic ratio with $\sigma $
requires an extra factor $1/\rho ^{2}$. We have the representation 
\begin{eqnarray}
&&\frac{1}{\rho ^{2}}\sigma _{\mathrm{el}}^{R}(s)=\frac{1}{\rho ^{2}}%
\int_{0}^{\infty }dt~T_{R}(s,t)^{2}~dt  \label{elastic_param} \\
&=&10.2037+2.47691\log \sqrt{s}+0.23108\log ^{2}\sqrt{s}~.  \notag
\end{eqnarray}%
The asymptotic ratio is now $(1/\rho ^{2})(\sigma _{\mathrm{el}}^{R}/\sigma
)\rightarrow 0.188$. These ratios participate in the geometric
interpretation in b-space representations.

\begin{figure*}[b]
\includegraphics[width=8cm]{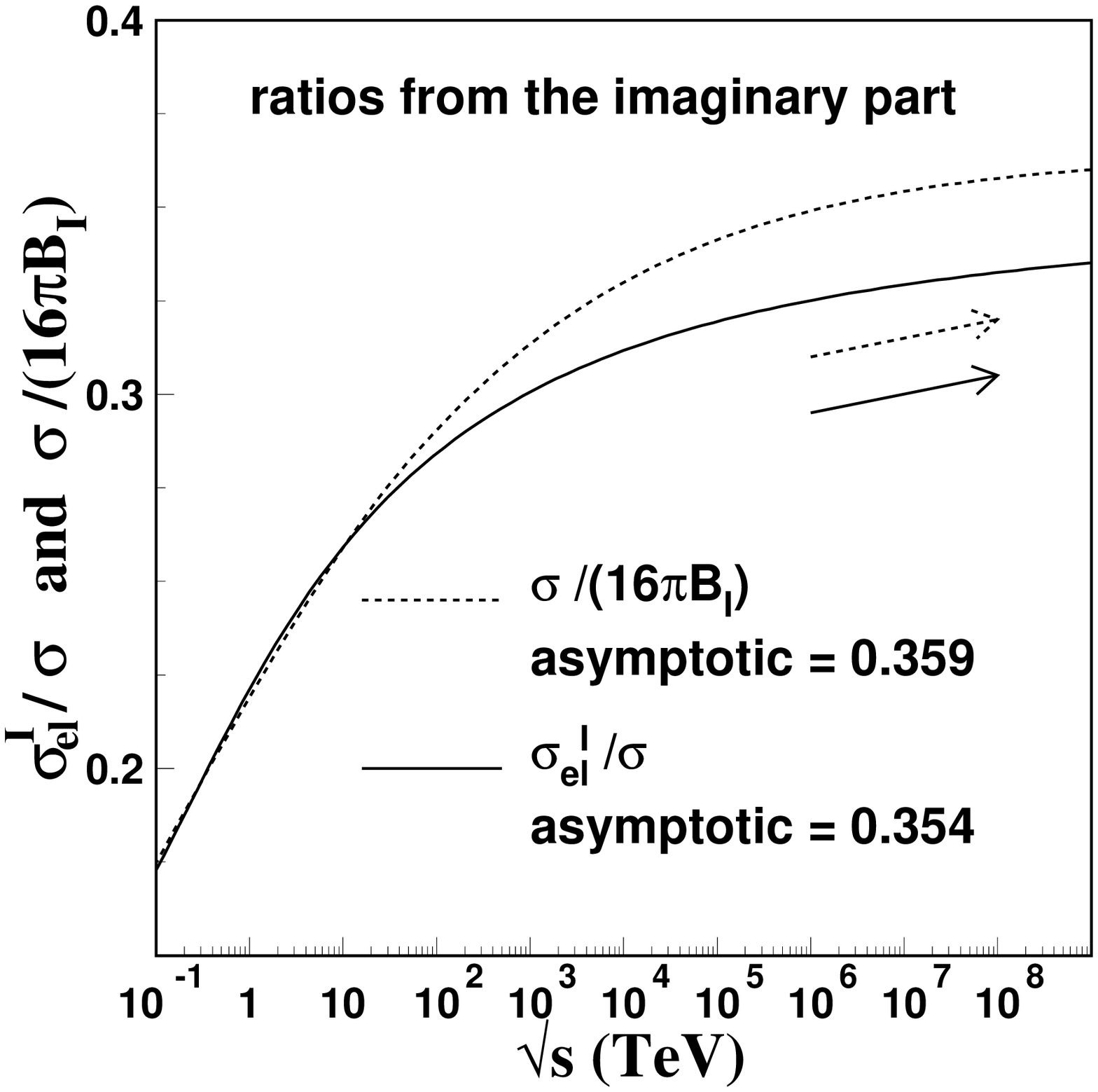} %
\includegraphics[width=8cm]{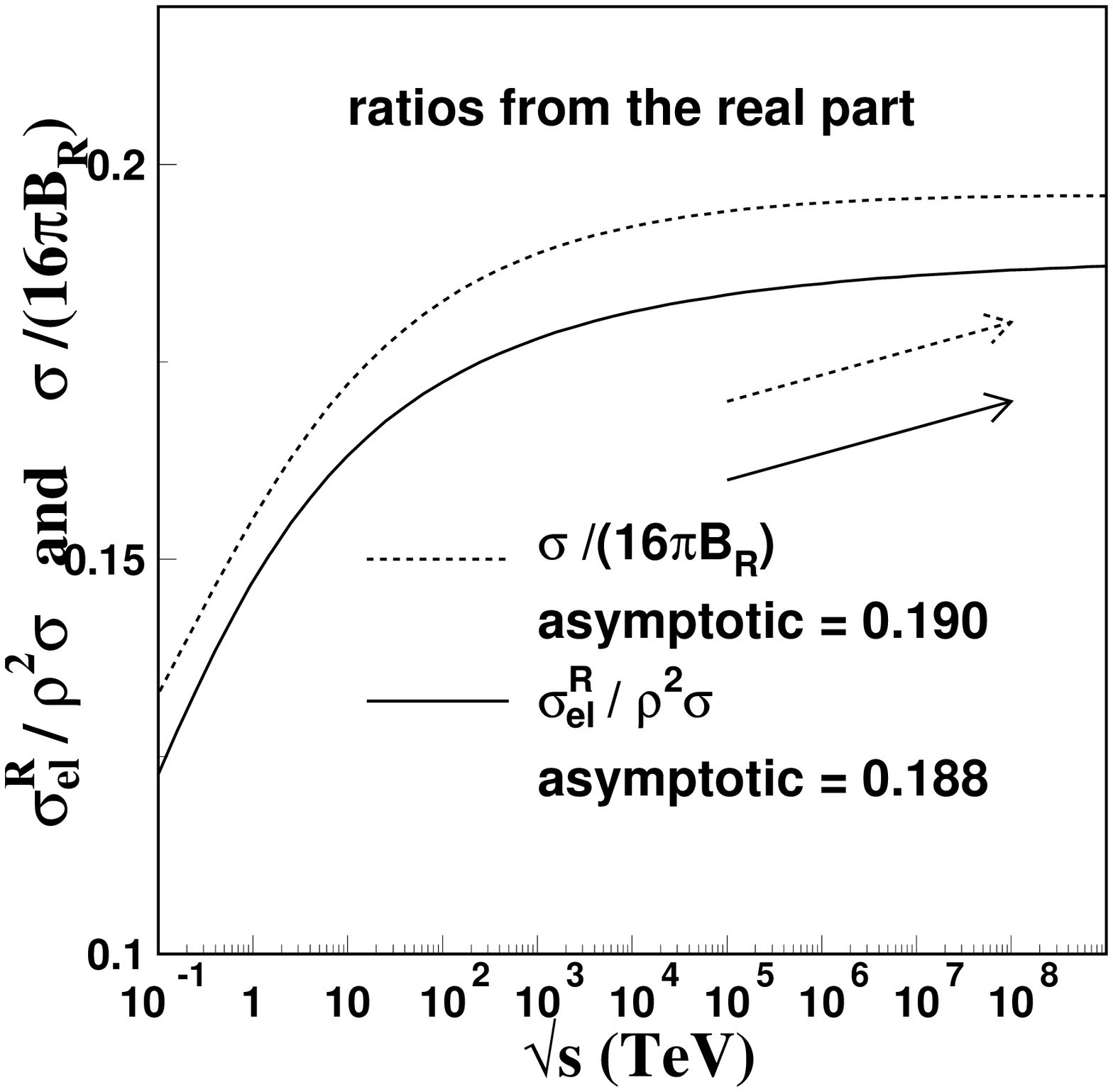} 
\caption{ Ratio between integrated (imaginary part) elastic cross section
and total cross section and ratio between total cross section and imaginary
slope as function of energy. On the RHS, the same for the real sector. The
asymptotic limits are approached very slowly : observe the extended energy
scale. For each part (Imaginary or Real) the two kinds of ratio would be the
equal if the amplitudes were of purely exponential form. We may observe that
the ratio of ratios in each sector (I or R) ir about the same, namely $%
0.359/0.354 \approx 0.190/0.188 \approx 1.01$. }
\label{ratios-fig}
\end{figure*}

The dimensionless ratios 
\begin{equation}
\sigma /(16\pi B_{K}),~~~~K=I,R
\end{equation}%
are related to $\sigma _{\mathrm{el}}^{I}/\sigma $ and $(1/\rho ^{2})\sigma
_{\mathrm{el}}^{R}/\sigma $ when the amplitudes are of pure exponential
forms with $B_{I}$ and $B_{R}$ slopes. The imaginary part is studied to
investigate the occurrence of black disk behaviour (assuming zero real
part), where the ratios $\sigma _{\mathrm{el}}^{I}/\sigma $ and $\sigma
/(16\pi B_{I})$ are both equal to 1/2. As shown in Fig. \ref{ratios-fig} our
solutions lead to values about 1/3 for the imaginary part case, which is a
more realistic expectation \cite{menon} than the black disk hypothesis.

\subsection{Geometric Scaling and Ratio of Cross Sections}

\begin{figure*}[b]
\includegraphics[width=8cm]{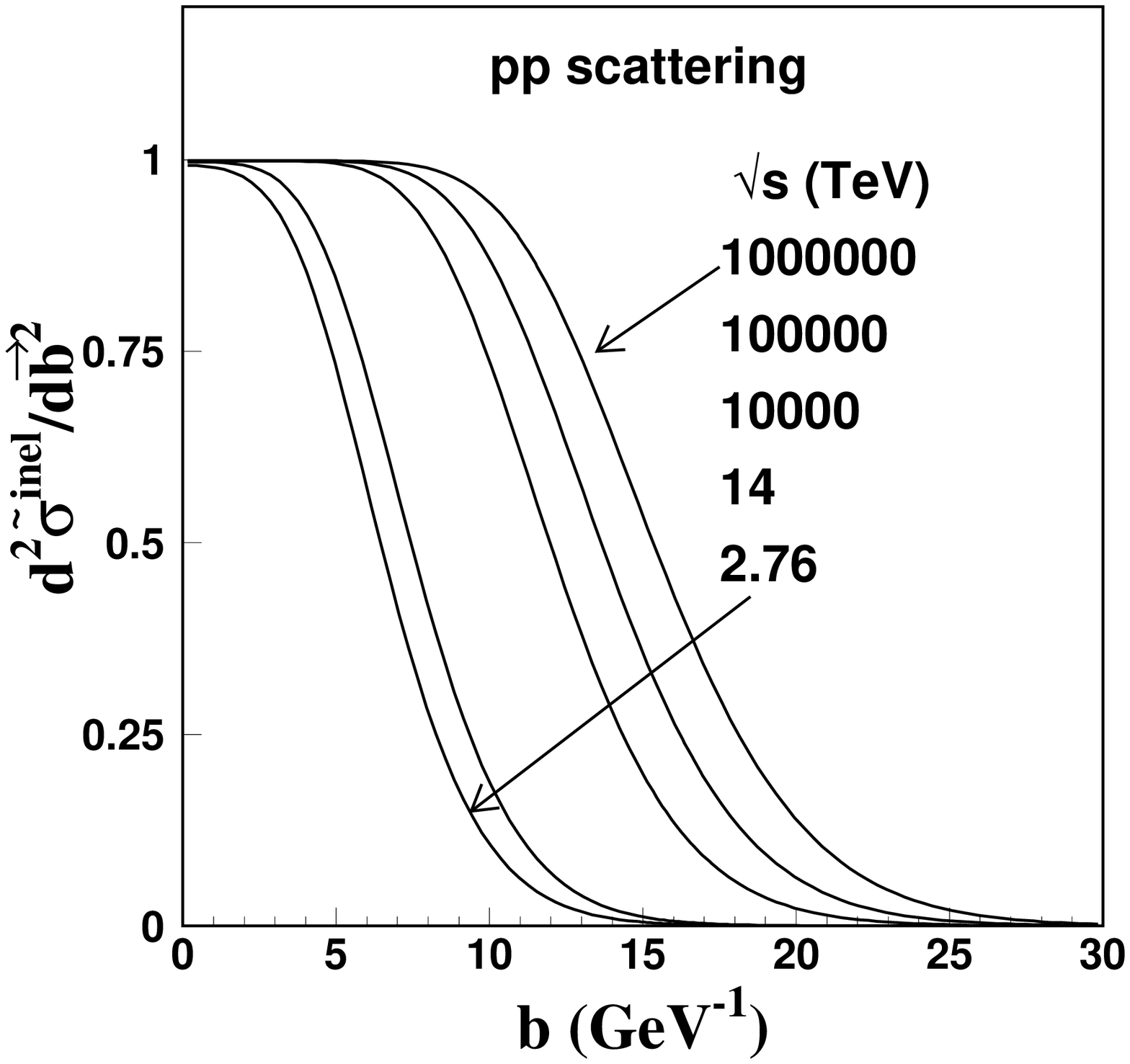} %
\includegraphics[width=8cm]{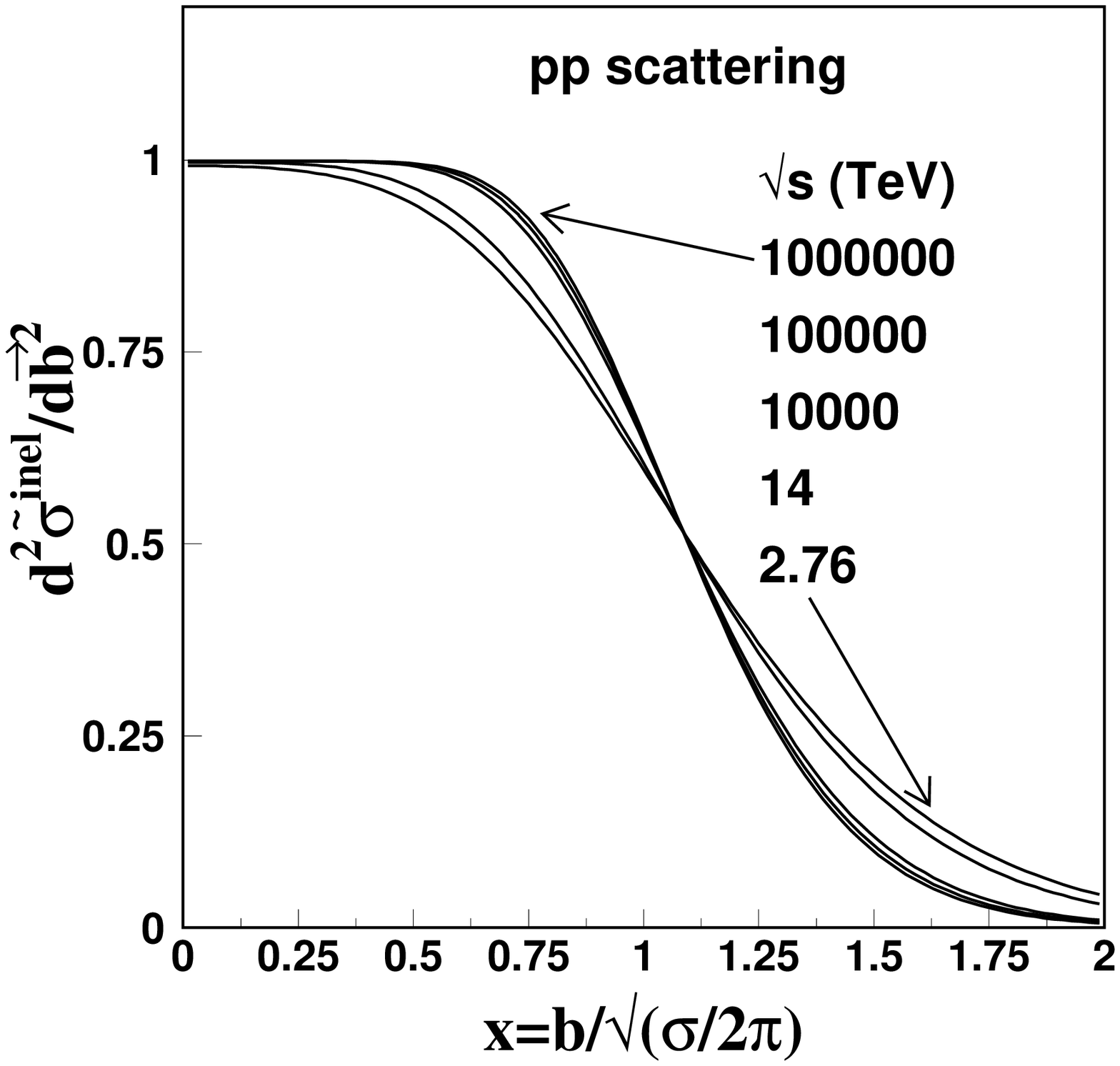}
\caption{ a) Plots of $d^{2}\protect\sigma _{\mathrm{inel}}/d\vec{b}^{2}$ as
function of $b$ for $\protect\sqrt{s}=$ 2.76, 14 TeV and for three very high
energies indicated in this figure; b) the same quantity plotted as function
of scaled variable $x=b/\protect\sqrt{\protect\sigma \left( s\right) /2%
\protect\pi }$, showing the convergence to a unique function, $\protect\xi %
\left( x\right) $ which has a finite surface diffuseness. }
\label{scaling-fig}
\end{figure*}

In Fig. \ref{scaling-fig}-a, we plot $d^{2}\sigma _{\mathrm{inel}}/d\vec{b}%
^{2}$ defined in Eq. (\ref{dsdb_inel}) as function of $b$ for $\sqrt{s}=$
2.76 and 14 TeV. The behavior at very high energies ( $\sqrt{s}=10^{4},10^{5}
$ and $10^{6}$ TeV) is also shown. We clearly see the increase of effective
radius of the interaction range with increasing energy. In Fig. \ref%
{scaling-fig}-b, we plot the same quantities with respect to the variable 
\begin{equation}
x\equiv \frac{b}{\sqrt{\sigma \left( s\right) /2\pi }}~.
\end{equation}%
This figure shows clearly that there exist a universal function $\xi \left(
x\right) $ such that%
\begin{equation}
d^{2}\sigma _{\mathrm{inel}}/d\vec{b}^{2}\rightarrow \xi \left( x\right) 
\end{equation}%
for $\sqrt{s}\gg 10^{4}$ TeV. An important point is that $\xi \left(
x\right) $ is far from the Heaviside step function, rather possesing a
considerably diffused surface. In this asymptotic limit, we can safely set $%
\cos \chi _{R}\rightarrow 1$ so that the total cross section is \cite%
{KEK-to-be} 
\begin{equation}
\frac{d^{2}\sigma \left( s,b\right) }{d\vec{b}^{2}}\rightarrow 2\left( 1-%
\sqrt{1-\xi (x)}\right) .~  \label{sigppp-tot-scale}
\end{equation}%
Note that $0\leq \xi \leq 1$ means $\left( 1-\xi \right) \leq \sqrt{1-\xi },$%
\ so that $\xi \left( x\right) \geq 1-\sqrt{1-\xi \left( x\right) }$ for all 
$x$ where the equality holds if and only if $\xi =0$ or $\xi =1$. Therefore,
whenever the function $\xi $ is different from a sharp-cut Heaviside theta
function $\theta \left( 1-x\right) ,\ $ we have%
\begin{equation}
\frac{\int_{0}^{\infty }x~\xi (x)~dx}{2\int_{0}^{\infty }x~\left( 1-\sqrt{%
1-\xi (x)}\right) ~dx}>\frac{1}{2}~.  \label{Constraint}
\end{equation}%
For our amplitudes, as shown in Fig. \ref{scaling-fig}, $\xi $ clearly does
not converge to a sharp-cut $\theta \ $function, preserving an appreciable
diffused surface for asymptotic energies. Therefore, we have 
\begin{equation}
\frac{\sigma _{\mathrm{inel}}}{\sigma \left( s\right) }>\frac{1}{2},
\end{equation}%
or%
\begin{equation}
\frac{\sigma _{\mathrm{el}}\left( s\right) }{\sigma \left( s\right) }<\frac{1%
}{2}.
\end{equation}%
This means that our amplitudes do not show the black disk behavior at very
large energies, deviating from the well known result for a black disk $%
\sigma _{\mathrm{el}}/\sigma \left( s\right) \rightarrow 1/2$. From the
above discussion, we can also easily see that the more diffused surface $\xi
\left( x\right) $ has, the less the ratio $\sigma _{\mathrm{el}}\left(
s\right) /\sigma \left( s\right) $ becomes. In fact, for our case this ratio
is close to $1/3$ (see also \cite{menon}). Note that this is somewhat
different scenario compared to \cite{deus}, where $\xi \left( x\right) $
would not have surface diffuseness.

\clearpage

\section{ Comparison with Data and Predictions  \label{data}}

Our description \cite{KEK_2013} of the elastic scattering data at 7 TeV from
the TOTEM Collaboration \cite{TOTEM_7} reproduces N=165 points in $d\sigma/dt
$ with an impressive squared average relative deviation $<\chi^2>=0.31$.
Characteristic quantities at this energy, shown in Tables \ref{first_table}
and \ref{second_table} are $\sigma=98.65$ mb, $\sigma_{\mathrm{el}}=25.39$
mb , $B=19.90$ GeV$^{-2}$, that compare extremely well with the values
published by TOTEM \cite{TOTEM_7}, $\sigma=98.6 \pm 2.2$ mb, $\sigma_{%
\mathrm{el}}=25.4 \pm 1.1$ mb , $B=19.9 \pm 0.3$ GeV$^{-2}$.

After the successful description of the 7 TeV data \cite{KEK_2013}, we now
present comparison and predictions for other LHC energies.

\subsection{ Inelastic and Total Cross Sections}

\begin{figure}[b]
\includegraphics[width=8cm]{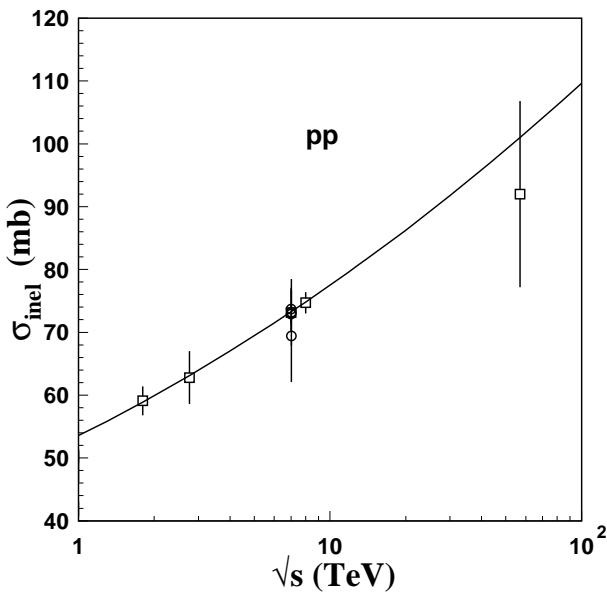} 
\caption{ Our calculations for pp inelastic cross sections and the data
above $\protect\sqrt{s}= 1 $ TeV , that cover the energies 1.8 TeV 
\protect\cite{Klimenko}, 2.76 TeV \protect\cite{ALICE}, 7 TeV \protect\cite%
{TOTEM_7, TOTEM_7b, TOTEM_7c,ALICE, ATLAS}, 8 TeV \cite{TOTEM_8} and 
57 TeV \protect\cite{Auger}. }
\label{inelastic-fig}
\end{figure}

For the inelastic cross section we assume the difference $\sigma_{\mathrm{%
inel}}=\sigma-\sigma_{\mathrm{el}}$ and then we have  73.26 mb at 7 TeV..
Published values of the TOTEM Coll. using different methods are $73.15 \pm
1.26$ \cite{TOTEM_7}, $73.7 \pm 3.4$ \cite{TOTEM_7b} and $72.9 \pm 1.5$ \cite%
{TOTEM_7c}. ALICE Coll. \cite{ALICE} gives  $\sigma_{\mathrm{inel}}= 73.2
\pm 5.3 $ mb , and ATLAS Coll. $\sigma_{\mathrm{inel}}= 69.4 \pm 2.4 \pm 6.9 
$ mb \cite{ATLAS}. We are not able to understand the CMS results \cite{CMS}
in terms of pure $\sigma_{\mathrm{inel}}$ due to non-informed missing
contributions. In these measurements there are extrapolations to using Monte
Carlo models to include diffractive events of low mass. Of course all these
results are compatible with our calculations.

A  measurement to be compared with our predictions is the $\sqrt{%
s}=2.76$ TeV value of ALICE Coll., that gives $\sigma_{\mathrm{inel}} = 62.8
\pm 4.2 $ mb , while our tables give the compatible value 63.11 mb.

The analysis of compatibility for the 1.8 TeV measurements of $\sigma_{%
\mathrm{inel}}$ by CDF and E811 in Fermilab  \cite{Klimenko} suggests the
value $ (1+\rho^2)\sigma_{\rm inel}=(60.3 \pm 2.3$ mb, that with with our $\rho$
value gives  $ \sigma_{\rm inel}=(59.1 \pm 2.3$ mb. Our table gives
58.89 mb for 1.8 TeV, once more in very good agreement.

Finally, at 57 TeV the Auger Cosmic Ray experiment \cite{Auger}, 
using other models for the pp input,  evaluates $%
\sigma_{\mathrm{inel}} = 92 \pm 14.8 $ mb , while our extrapolation gives
101 mb. We have discussed this measurement \cite{KEK_CR_2014} together with
other CR Extended Air Showers (EAS) experiments, using our amplitudes as
inputs and a basic Glauber method to connect pp and p-air processes. Our
calculation reproduces well all CR data for p-air cross sections  
with $\sqrt{s}$ (in the pp system)  up to 100 TeV.   
 
For 8 TeV we have predictions $\sigma=101.00$ mb , $\sigma_{\mathrm{el}%
}=26.18$ mb , $\sigma_{\mathrm{inel}}=74.82$ mb , $\sigma_{\mathrm{el}%
}/\sigma=0.26 $ shown in the tables. The measurements by TOTEM \cite{TOTEM_8}
give for the same quantities $\sigma=101.7 \pm 2.9 $ mb , $\sigma_{\mathrm{el%
}}=27.1 \pm 1.4 $ mb, $\sigma_{\mathrm{inel}}=74.7 \pm 1.7$ mb , $\sigma_{%
\mathrm{el}}/\sigma=0.266\pm0.006$. Of course these numbers are very
encouraging, indicating also good expectations for $d\sigma/dt$ at this energy. 

The data and our curve for $\sigma_{\mathrm{inel}}(s)$ are shown in Fig. \ref%
{inelastic-fig}.  All this information shows that our formulae for the
energy dependence of $\sigma(s)$ and $\sigma_{\mathrm{inel}}(s)$ in pp 
scattering work very well.

\subsection{Expected data for $d\protect\sigma /dt$ at 8 TeV \label%
{dsigdt_8TeV}}

The  preliminary   data for $d\sigma/dt$ at 8 TeV,  
shown in talks by members of the TOTEM Collaboration \cite{TOTEM_talks}, 
are encouraging for the application of our method of analysis. 
We recall that in the treatment of the 7 TeV data, we obtained 
precise  description, with average $\langle \chi^2\rangle = 0.34 $ for 
165 data points in the whole $|t|$ interval of measurements.  

\begin{figure}[b]
\includegraphics[width=8cm]{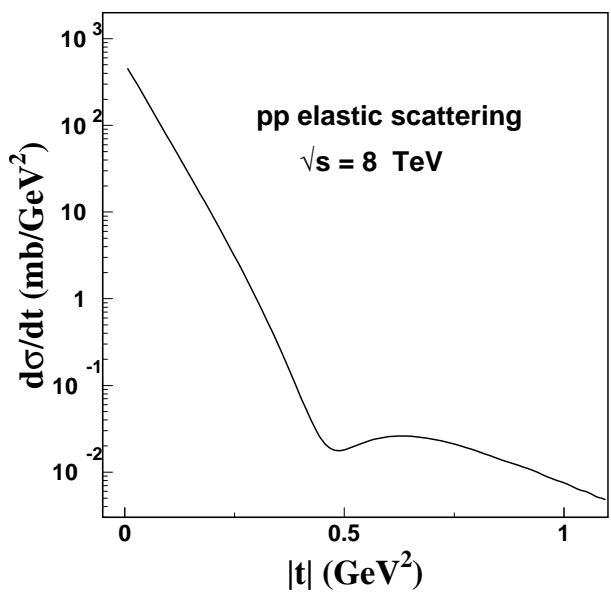} 
\caption{ Predicted representation for $d\protect\sigma/dt$ in the 
whole $|t|$ range of observations at 8 TeV made in LHC by the  
TOTEM Collaboration \cite{TOTEM_talks}.  }
\label{data_8TeV-fig}
\end{figure}

If Fig. \ref{data_8TeV-fig} we shown our calculation for $d\sigma/dt$ 
covering the whole $|t|$ range of the preliminary information, 
using the amplitudes defined in Sec. \ref{energy_range}. The 
characteristic features of the forward peak and of the dip/bump 
structure are expected to represent accurately the angular 
dependence. Numerical values for  characteristic features are given 
in   Tables \ref{first_table} and \ref{second_table}. 

This is the description of the global $d\sigma/dt$ data at 8 TeV, that 
promises to be more complete and regular  than the 7 TeV
data, except for not reaching larger $|t|$ values. 
In the following we discuss the forward region in more detail.

\begin{figure*}[b]
\includegraphics[width=8cm]{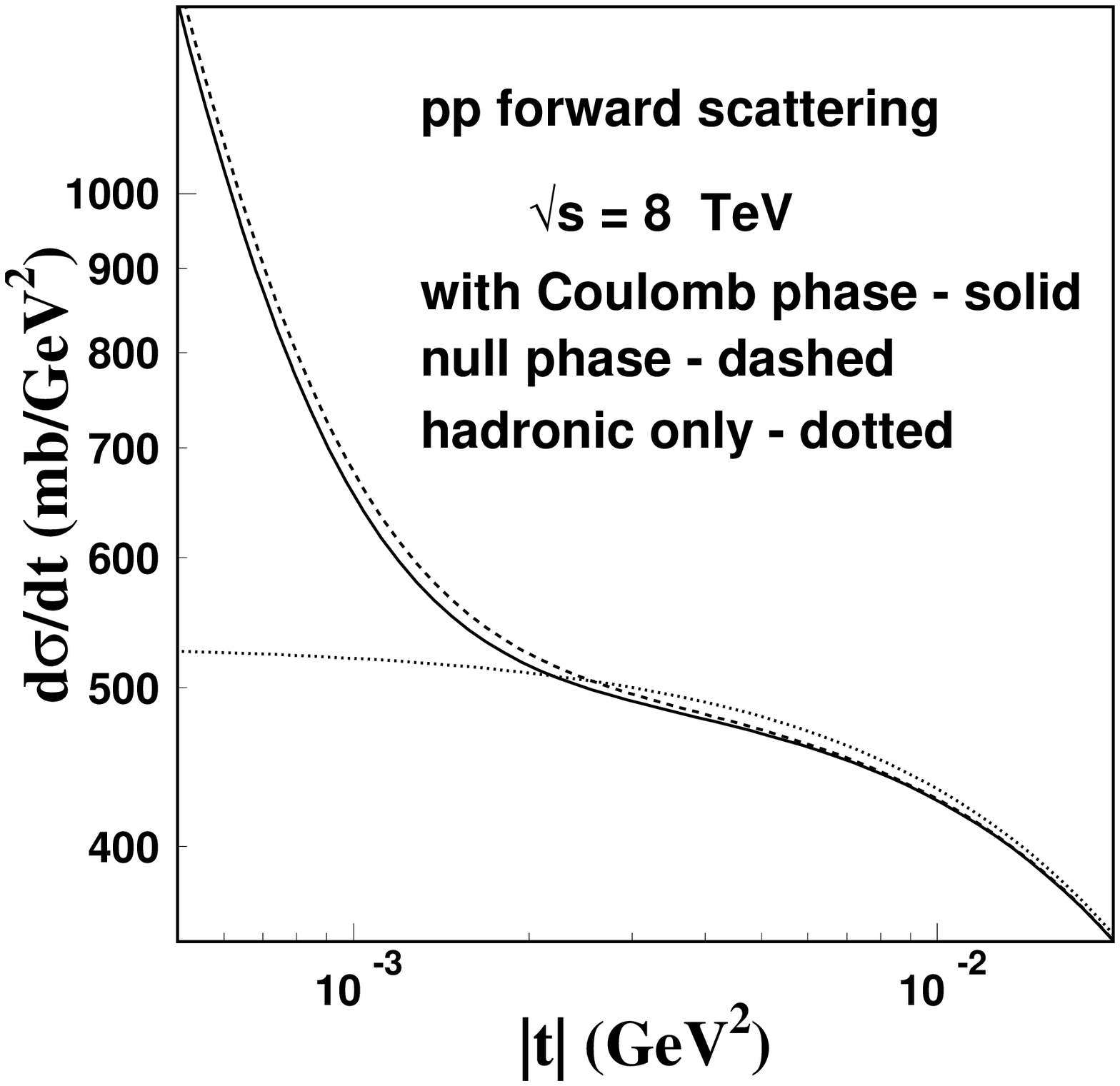} %
\includegraphics[width=8cm]{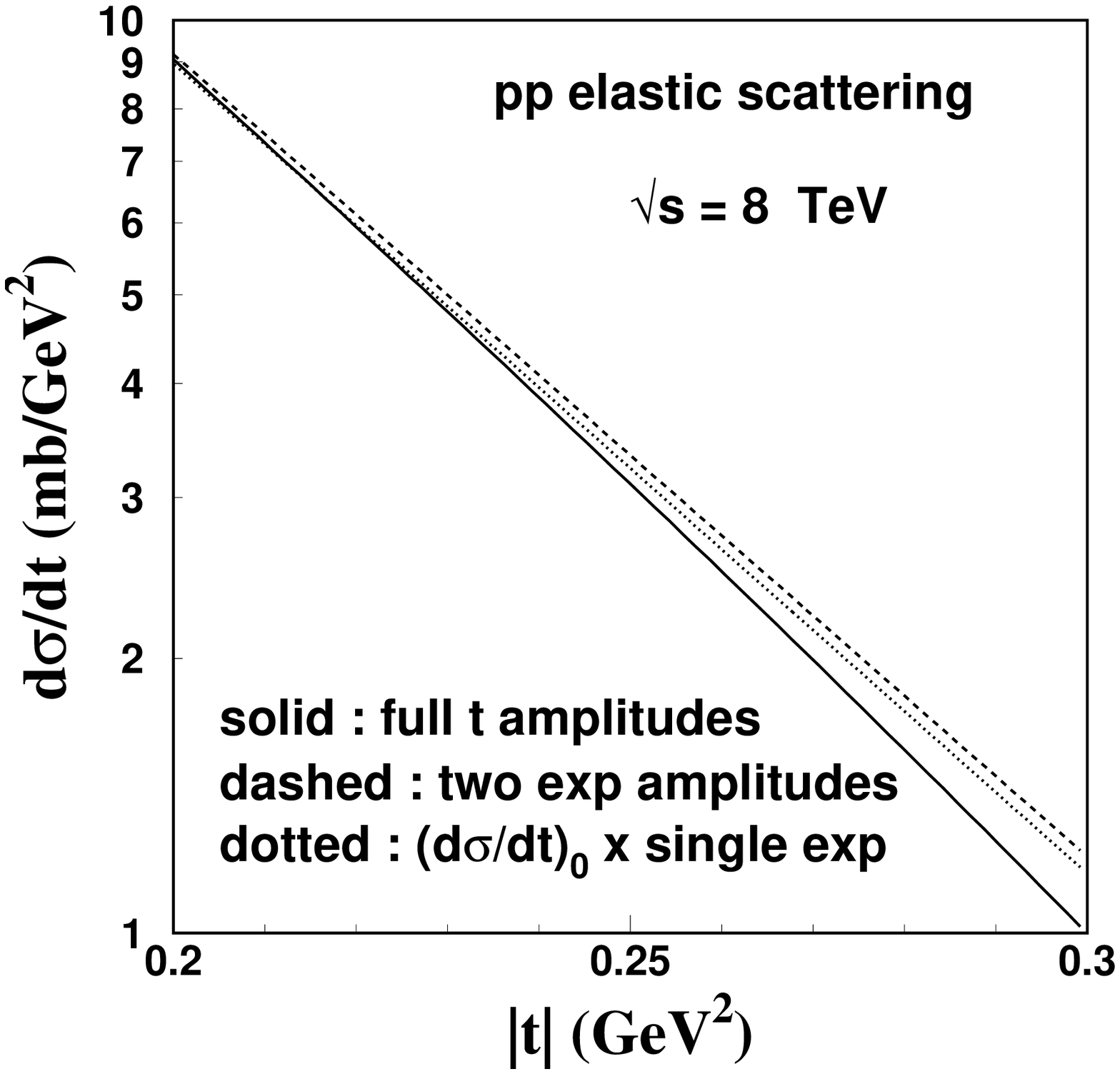}  
\caption{ Calculation of $d\sigma/dt$ in the  forward range at 8 TeV.
The solid lines correspond  to the full calculation with our amplitudes. 
a) In the LHS  a $\log |t|$ scale is used to represent in detail the 
forward range; the dashed line is obtained with Coulomb phase put equal to 
zero; the dotted line represents the hadronic interaction. 
b) In the RHS The dashed line represents the calculation with pure exponential 
amplitudes, with the real and imaginary parts entering with their 
corresponding slopes; the dotted line shows the usual description of 
the foreword peak in the form 
$d\sigma/dt= d\sigma/dt|_{t=0} ~  \exp{(-B|t|)}$ ~ . }
\label{forward_8TeV-fig}
\end{figure*}

In Fig. \ref{forward_8TeV-fig} we plot the calculations in the small 
$|t| $ range , including  the influence of the
Coulomb phase \cite{KEK_2013}. The calculation with Coulomb phase put equal
to zero is represented by the dashed line, showing that its influence is
small. Our specific calculation of the Coulomb phase takes into account 
the difference in values of the $B_R$  and $B_I$ slopes.  
Other calculations for the interference phase \cite{phases} also show
that its influence is small, reducing $d\sigma/dt$ by a few percent. 
  
Our values for $B_I$ and $B_R$ given in Table \ref{first_table} lead to the $%
d\sigma/dt$ effective slope at 8 TeV 
\begin{equation}
B=\frac{B_I+\rho^2 B_R}{1+\rho^2}
\end{equation}
equal to $B=20.405 ~ \nobreak\,\mbox{GeV}^{-2}$.

Our predictions seem to be in accordance with the eye-guided reading of 
the preliminary data of $d\sigma/dt$ that appear  in presentations of 
the TOTEM group in workshops,  at least at the qualitative level.
At 7 TeV our expressions perform extremely well when
compared to the published experimental information, and we expect that the
same will happen at 8, 13 and 14 TeV .

  \subsection{Other models\label{other_models-sec}}

The complete identification of the real and imaginary parts 
of the complex  pp elastic amplitude  is fundamental for the 
knowledge of the dynamics of the  collision, being 
an essential  bridge between the observed quantities and fundamental 
QCD dynamical processes. Our determination, though consistent 
and complete, depends on the analytical forms used for  the 
representation. It is thus important to compare our predictions 
with the results obtained with other input assumptions.   

The important Yukawa-like behaviour of the amplitudes in $b$-space 
based on the behaviour of  the loop-loop interaction for large $b$ 
in the Stochastic Vacuum Model,  that is incorporated in our input 
amplitudes in Eq. (\ref{Shape-b}), is confirmed  in a recent 
treatment of the pp interaction through Wilson correlation functions 
\cite{Matteo}.

  The representation of amplitudes in $b$-space  from the ISR to the SPS 
energies shows  at $b=0$ a slow increase with the energy  \cite{boris1},  
remaining below  saturation, which   seems to be approached asymptotically, 
as can be seen in the present work  for the LHC energies and also 
in studies at higher cosmic ray energies \cite{CR_2014}.  
With parameters adjusted to describe the energy dependence 
in the 23 GeV - 546 GeV range \cite{boris2} the model predicts  
characteristic quantities of pp forward scattering  for the range
7 - 14 TeV, with results for the total cross section and the slope 
parameter that  agree very well with the  numbers given in 
Table \ref{first_table}, and in particular with the TOTEM values at 7 TeV.  

\begin{figure*}[b]
\includegraphics[width=8cm]{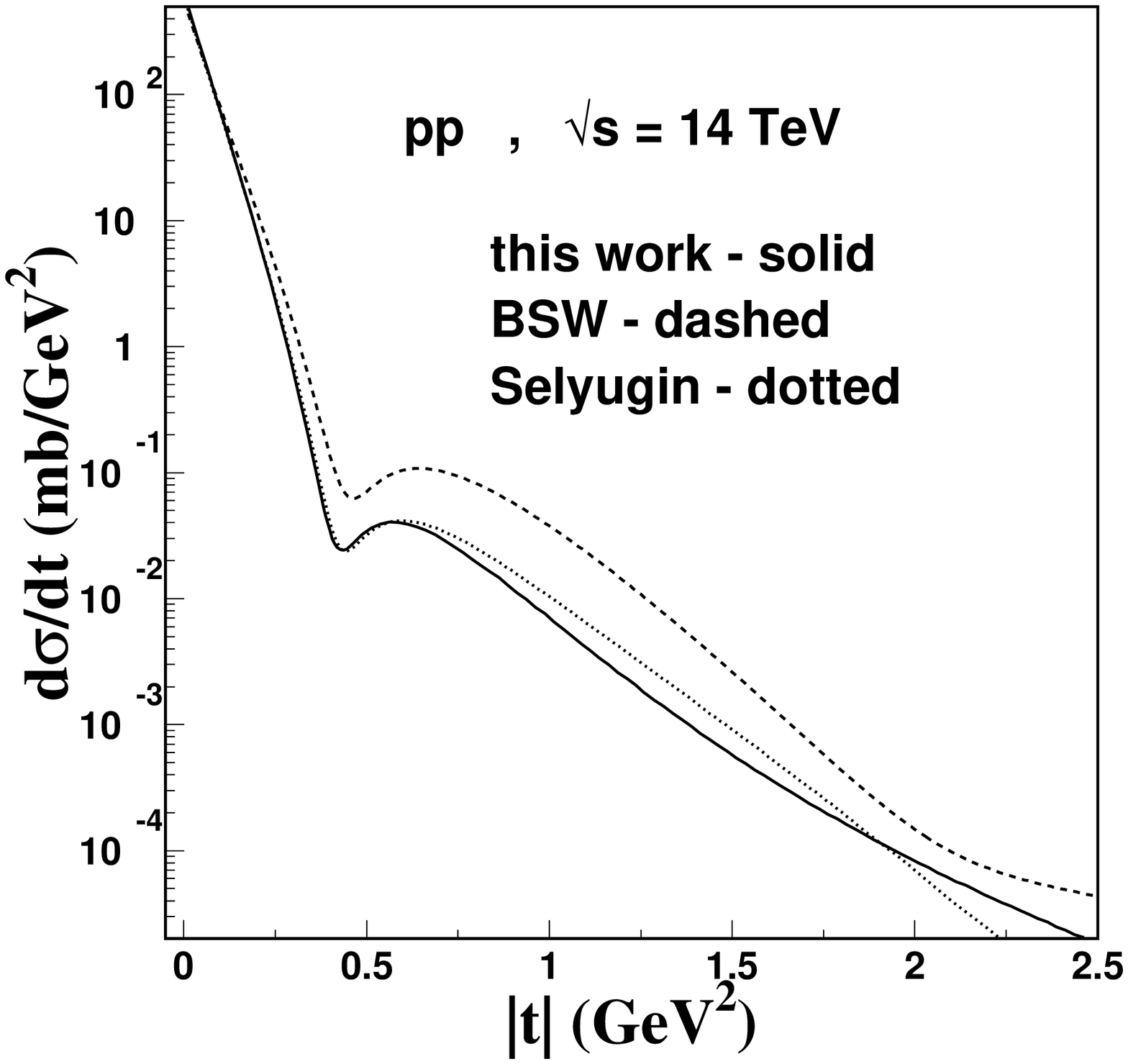} %
\includegraphics[width=8cm]{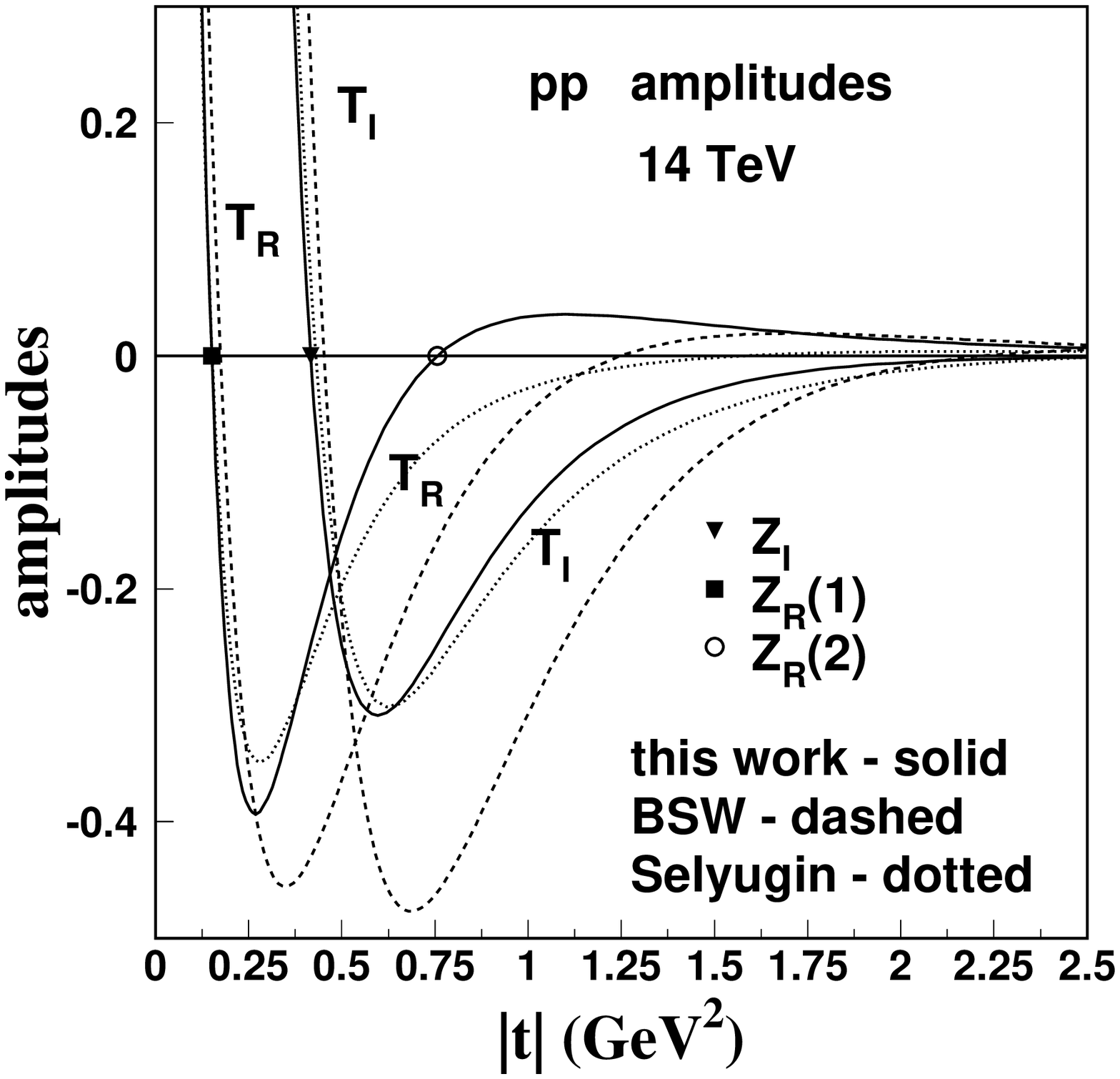}  
\caption{ Calculation of $d\sigma/dt$ at 14 TeV compared with 
the predictions of the BSW model.
The solid lines correspond  to the full calculations with our amplitudes. 
a) In the LHS are shown the differential cross sections. 
b) In the RHS are shown  the real and imaginary parts of the amplitudes. }
\label{BSW-fig}
\end{figure*}

The model proposed by Bourrely, Soffer and Wu (hereafter called BSW model) 
 \cite{BSW} gives explicitly the full s,t dependence of the elastic 
scattering amplitudes and is appropriate for the comparison with our 
results. Important similarities and differences were discussed in detail 
in the 7 TeV case \cite{KEK_2013}, and we now compare the predictions for 
14 TeV.  Fig. \ref{BSW-fig} shows that the dip-bump structure occurs in  
similar $|t|$ regions, but there is a difference in $d\sigma/dt$ by a factor 
larger than 2. This difference results from the larger magnitudes (with 
negative signs) of both real and imaginary parts in the BSW model, 
as can be observed in the second part of the figure. The second real 
zero occurs for a larger $|t|$ in the BSW calculation.   
    
\begin{figure}[b]
\includegraphics[width=8cm] {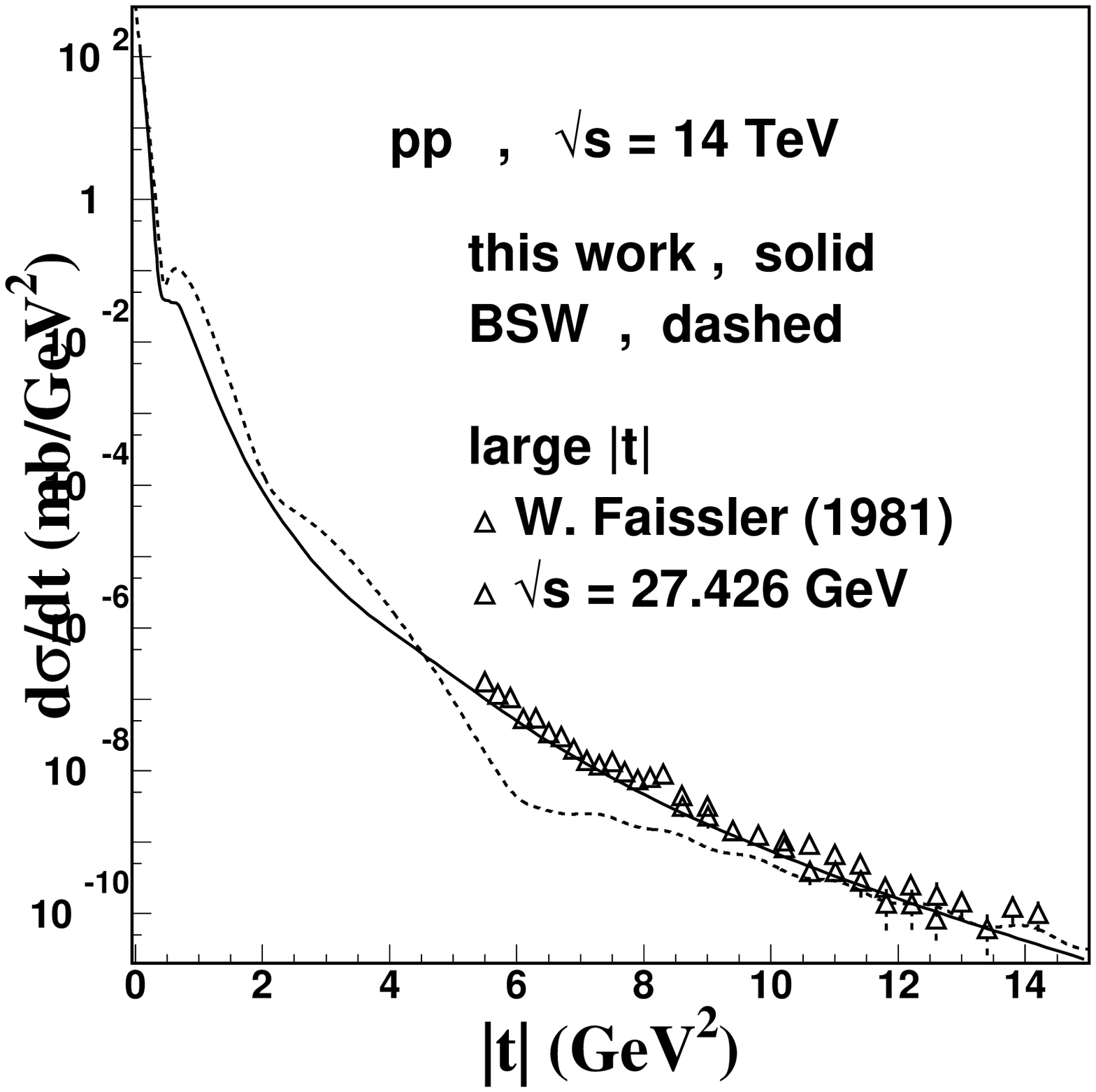}  
\caption{ Calculations  predicting $d\protect\sigma/dt$  
for large $|t|$ at 14 TeV,  including the perturbative tail term   
in our real amplitude, plotted   with  the data 
\protect\cite{Faissler} at 27.4 GeV for very large $|t|$. }
\label{tail-fig}
\end{figure}

To raise interest on   measurements at higher $|t|$, in 
 Fig. \ref{tail-fig} the 14 TeV plot is extended to very large 
$|t| $   pointing  out the possible smooth connection 
with the supposedly universal tail at 27.4 GeV \cite{Faissler}. 
We recall the situation with similar plot drawn in the 7 TeV 
case \cite{KEK_2013}, where the measurements reached larger 
$|t| \approx 2.5 \GeV ^2$ and the conjecture of the universality of 
the tail at such high energy encounters  motivation.  
The broad dip in the  region of 6 GeV$^2$ in the BSW calculation is 
due to a zero  in its  imaginary amplitude, as was also indicated 
at  7 TeV .

The structure of the pp and ${\rm p\bar p }$ interactions  
studied by O. Selyugin \cite{Selyugin}, based on the analysis  of 
different sets of Parton Distribution Functions and introducing 
t-dependence in the Generalized Parton distributions,  
 gives good representation of $d\sigma/dt$ data in large 
energy range, up to the LHC Totem experiment at 7 TeV.  
We include in Fig. \ref{BSW-fig} the $t $ dependences of 
$d\sigma/dt$ and  amplitudes in this model for 14 TeV. It is important 
to observe  the similarity with our results in the forms of the 
amplitudes, that differ essentially only in the real part for large $|t|$, 
with different locations of the second zero. The similarity, that must 
be investigated at all energies, reinforces the expectation of the 
present work, that is to find a realistic and accurate disentanglement 
of the elastic amplitudes. 

Recently, T. Cs\"{o}rg\"{o} et al. applied the multiple diffraction 
calculation of the Glauber-Valesco Model \cite{GV} to the TOTEM results for 
7 TeV and fitted the parameters of partonic charge form factor, obtaining 
good representation for the scattering cross section, including the 
dip-bump structure \cite{Tamas}. Parametrizing the partonic distribution 
function for each energy, the Glauber-Valesco model also describes the 
lower energy data from ISR and FNAL. The model leads to the conclusion 
that at LHC energies the pp interaction is not that of sharp-edged black-discs, 
but presents a considerable tail in the profile function \cite{Tamas2}. These 
observations are essentially the same as in predictions of our  
work \cite{KEK_2013, KEK-to-be, CR_2014}.

In addition, more detailed measurements of inelastic pp cross section 
in LHC experiments \cite{LHC-inel}, especially of diffractive-dissociation 
processes (DD), add very important information on the dynamics of the 
interaction,  as discussed by Lipari and Lusignoli \cite{Lipari}.
These studies and ours are complementary in the sense that while our
approach is based on the field theoretical model of interaction 
between two geometrical objects, their approach applies the multiple
diffractive model and introduces parametrizations of the partonic form
factor of the proton. A detailed comparison of the two approaches, 
together with the identification of the presence of diffraction in the 
inelastic part in our formalism, will give more precise insight on the 
physical roles behind the features of our description. Studies in this
direction are under investigation.

\section{\ Final Remarks and Comments}

In this paper we present  predictions for observables of pp elastic
scattering above $\sqrt{s}=1$ TeV  up to coming LHC energies in terms 
of analytic forms for the real and imaginary parts of the 
complex scattering amplitude. 
The representation proposes a separate identification of the  two parts,
which are both constructed respecting   unitarity and dispersion 
relation constraints, and precisely determine their  influences in  the 
observed quantities.  

The amplitudes have simple analytical forms, that can be directly 
evaluated with few  operations with elementary functions.
The shape of the dip-bump behavior results from a  delicate interplay of the
imaginary and real amplitudes. All intervening quantities and 
  derived  properties are connected by smooth energy dependences.

The zeros of the real and imaginary parts have very regular 
displacements, converging to   finite limits as the energy 
increases. There is remarkable connection 
between positions of zeros  and positions and heights of dips 
and bumps and   inflections in $d\sigma/dt$. 

The slopes $B_I$ and $B_R$ at the origin, with their characteristic 
difference in values, together with the ratio $\rho$,  are essential 
quantities that participate in the definition, through the unique analytical 
forms of the amplitudes, of the properties of the observed $d\sigma/dt$  
in the whole $t$ range. 
Their values are thus fixed with high accuracy. It is very 
important that the slopes show quadratic dependence in $\log{s}$,
instead of the linear dependence suggested by Regge phenomenology. 

The integrated elastic cross sections are evaluated  in their 
separate parts, obtained from the real and imaginary amplitudes, and 
are also represented by simple parabolic forms in $\log{s}$. 

The properties of ratios (with respect to the total cross section)  
of slopes and of integrated elastic cross sections,  that 
tend to finite asymptotic  limits, are studied, showing that the 
hypothesis of a black disk limit in the behaviour of the pp 
interaction  seems to be excluded by phenomenology.  

In Sec. \ref{data} we give  predictions, presented in tables 
\ref{first_table} and \ref{second_table}, in equations and in figures. 
Taking into account previous publications at 1.8 and 7 TeV, 
the present paper give explicit predictions of cross sections at 
2.76, 8 , 13 and 14 TeV, with no free  numbers.  
More precise future data   may  confirm our predictions more firmly.
 
We also discuss the geometrical interpretation of our amplitudes, showing 
that the effective interaction radius in $b$-space increases with the energy.
 Our amplitudes obey a geometric scaling in asymptotic energies, and
indicate that the profile function $d^{2}\sigma _{\mathrm{inel}}/d^{2}\vec{b}
$ tends to a universal (energy independent) function with respect to a
scaling variable, $x\sim b/\sqrt{\sigma }$.  This universal function exhibits 
a considerable diffused surface, indicating a   scenario different from 
the commonly accepted  black disk. At LHC energies, the saturation seems to start
(the central value of $d^{2}\sigma _{\mathrm{inel}}/d^{2}\vec{b}$ is almost
unity), but the asymptotic profile is still far and only  can be reached for 
$\sqrt{s}>10^{4} $ TeV. The connection between the diffused surface of long 
range and inelastic diffractive processes will be an interesting line of 
investigation.

We believe that our analytic representation of the
scattering amplitudes will serve as important guidance, not only for the
future measurements in LHC,  but also for a theoretical understanding
 of the intermediate region of partonic saturation phenomena.

\bigskip

\begin{acknowledgments}
The authors wish to thank the Brazilian agencies CNPq, PRONEX , CAPES and FAPERJ for
financial support.
\end{acknowledgments}


\begin{thebibliography}{99}
\bibitem{ferreira1} E.~ Ferreira and F.~ Pereira, \emph{Phys. Rev. D } 
\textbf{59} , 014008 (1998) ; \emph{Phys. Rev. D } \textbf{61}, 077507
(2000).
\bibitem{KEK_2013} A. Kendi Kohara, E. Ferreira and T. Kodama , \emph{Eur.
Phys. J. C} ,\textbf{73}, 2326 (2013). 
\bibitem{TOTEM_7} G. Antchev et al., Totem Coll., Eur. Phys. Lett. \textbf{%
101}, 21002 (2013)
\bibitem{KEK-to-be} A. K. Kohara , E. Ferreira and T. Kodama , "\textit{%
Energy and Asymptotic Behavior of pp scattering Amplitudes}", to be
published (2014).
\bibitem{CR_2014} A. Kendi Kohara, E. Ferreira and T. Kodama , Jour. Phys. G 
\textbf{41} (2014)115003 ; arXiv hep-ph 1406.5773.
\bibitem{dosch} H.G. Dosch, \emph{Phys. Lett. B } \textbf{190}, 177 (1987) ;
H.G. Dosch, E. Ferreira, A. Kramer \emph{Phys. Rev. D} \textbf{50}, 1992
(1994).
\bibitem{Regge} S. Donnachie, G. Dosch, P. Landshoff, O. Nachtmann , \textit{%
Pomeron Physics and QCD}, Cambridge University Press, 2002.
\bibitem{DL} A. Donnachie and P.V. Landshoff, Z. Phys. C \textbf{2}, 55
(1979); Phys. Lett. B \textbf{387}, 637 (1996).
\bibitem{KEK_2013b} A. Kendi Kohara, E. Ferreira and T. Kodama , Phys. Rev. 
D \textbf{87}, 054024, (2013).
\bibitem{ferreira2} E. Ferreira, Int. Jour. Mod. Phys. E \textbf{16}, 2893, (2007).
\bibitem{Martin} A. Martin, Phys. Lett. B \textbf{404}, 137 (1997).
\bibitem{menon} D.A. Fagundes, M.J. Menon, P.V.R.G. Silva, J. Phys. G40
(2013) 065005 ; ibid :  arXiv : 1410-4423[hep-ph] . 
\bibitem{deus} P. Brogueira and J. Dias de Deus, Jour. Phys. G \textbf{39}%
(2012) 055006 ; I. Bautista and J. Dias de Deus, Phys. Lett. B \textbf{718}
(2013), 1571. 
\bibitem{TOTEM_7b} G. Antchev et al., Totem Coll., Eur. Phys. Lett. \textbf{%
101} (2013) 21003.
\bibitem{TOTEM_7c} G. Antchev et al., Totem Coll., Eur. Phys. Lett. \textbf{%
101} (2013) 21004.

\bibitem{ALICE} B. Abelev et al., ALICE Coll., Eur. Phys. J. C (2013)
73:2456.

\bibitem{ATLAS} G. Aad et al., Nature Commun. 2:463 (2011).

\bibitem{CMS} S. Chatrchyan et al., CMS Coll., Phys Lett B \textbf{722}
(2013) 5.

\bibitem{Klimenko} S. Klimenko, J. Konigsberg and T. M. Liss, 
FERMILAB-FN-0741 (2013).

\bibitem{Auger} P. Abreu et al , Auger Coll., Phys. Rev. Lett. \textbf{109},
062002 (2012).

\bibitem{KEK_CR_2014} A. Kendi Kohara, E. Ferreira and T. Kodama , ArxiV 
hep-ph 1406.5773.

\bibitem{TOTEM_8} G. Antchev et al., Totem Coll., Phys. Rev. Lett. \textbf{%
111} (2013) 012001

\bibitem{TOTEM_talks} M. Deile, Totem Coll., Talk at DIS 2014 (Warsaw, April
2014); J. Kaspar, Totem Coll., talk at XXX-th International Workshop on High
Energy Physics, Protvino, June 2014.

\bibitem{phases} V. Kundr\'at and M. Lokaj\'icek, Phys. Lett. B \textbf{611}
(2005) 102 ; R. Cahn, Z. Phys. C \textbf{15} (1982) 253.

\bibitem{Matteo} M. Giordano and E.Meggiolaro, Jour. High Energy Phys. 03 (2014) 002.

\bibitem{boris1}  B.Z. Kopeliovich, I.K. Potashnikova, B. Povh and E. Predazzi, 
Phys. Rev. Lett. {\bf 85}, 507 (2000) ; ibd. Phys. Rev. D {\bf 63}, 054001 (2001). 
 
\bibitem{boris2} B.Z. Kopeliovich, I.K. Potashnikova and B. Povh, 
       Phys. Rev. D {\bf 86}, 051502 (2012). 
\bibitem{BSW} C. Bourrely, J.M. Myers, J.Soffer and T.T. Wu , 
     \emph{Phys. Rev.D } \textbf{85}, 096009 (2012).
\bibitem{Faissler} W. Faissler et al., Phys. Rev. D \textbf{23} , 33 (1981).
\bibitem{Selyugin} O.V. Selyugin,  \emph{Eur. Phys. J. C}(2012), 72:2073 ; 
talk  presented at Diffraction 2014 ; private communication is gratefully 
acknowledged. 
\bibitem{GV} R.J. Glauber and J.Velasco, Phys. Lett. B147 (1984) 380
\bibitem{Tamas} T.Csorgo, R. J. Glauber, F. Nemes,  arXiv:1311.2308v1
[hep-ph] (2013)
\bibitem{Tamas2} F. Nemes and T. Cs\H{o}g\H{o}, arXiv:1204.5617v2 [hep-ph]
(2012), T. Cs\H{o}g\H{o}, talk presented at International Workshop on
Collectivity in Relativistic Heavy Ion Collisions, Kolymbari, Crete, Greece,
Sept 14-20, 2014. 
\bibitem{LHC-inel} B, Abelev \textit{et al}. [The ALICE Collaboration],
arXiv:1208.4968[hep-ex]; G, Aad \textit{et al}. [ATLAS Collaboration],
Nature Commun, 2, 463(2011) [arXiv:11042.0326]] ; S, Cgatrchyan \textit{et al%
}. [CMS Collaboration], Phys.Lett.B 722, 5 (2013).
\bibitem{Lipari} P. Lipari and M. Lusignoli,  
\emph{Eur. Phys. J. C}(2013), 73:2630. 

\end{thebibliography}
\end{document}